\documentclass[11pt]{article}
\usepackage{cite}
\usepackage{amsfonts} 
 \usepackage{amssymb}
\usepackage[vcentermath]{youngtab}
 \def\bsh{\backslash}

 \newfont{\bbbold}{msbm10}
 \newfont{\cals}{eusm9}
\def\half{{1 \over 2}}
\def\com{\mbox{\bbbold C}}
 
\def\int{\mbox{\bbbold Z}}

\def\bb1{\mbox{\bbbold 1}}

 \def\cE{{\cal E}}
 \def\cF{{\cal F}}

 \def\cI{{\cal I}}

 \def\cO{{\cal O}}

 \def\cR{{\cal R}}
 \def\cS{{\cal S}}
 \def\cT{{\cal T}}
 
 \def\cV{{\cal V}}

 \def\cY{{\cal Y}}
 
 \def\co{\mbox{\cals O}}
\def\ct{\mbox{\cals T}}

 \newfont{\goth}{eufm10 scaled \magstep1}

 \def\gg{\mbox{\goth g}}

 \def\gl{\mbox{\goth l}}

 \def\go{\mbox{\goth o}}
 \def\gp{\mbox{\goth p}}

 \def\gs{\mbox{\goth s}}

 \def\a{\alpha} \def\A{\alpha}
 \def\b{\beta} \def\B{\beta}
 \def\c{\gamma}\def\C{\Gamma}
 \def\d{\delta}\def\D{\Delta}
 \def\e{\epsilon}
 \def\F{\Phi}\def\vf{\varphi}

 \def\l{\lambda}
 \def\m{\mu}

 \def\th{\theta}
 
 \def\be{\begin{equation}}\def\ee{\end{equation}}
 \def\bea{\begin{eqnarray}}\def\eea{\end{eqnarray}}
 \def\ba{\begin{array}}\def\ea{\end{array}}

 \def\del{\partial}
 \def\ua{\underline{\alpha}}
 \def\ub{\underline{\phantom{\alpha}}\!\!\!\beta}

 \def\unA{\underline A}
 \def\unB{\underline B}
 \def\unC{\underline C}

 \def\sd{\rm sdet}\def\str{\rm str}
  \def\sdet{{\rm sdet}}
\def\xz{\times}
 \def\cxz{\circ \hspace{-7.5pt} \diagup \hspace{-10pt}  \diagdown}
 \def\btx{\bt \hspace{-7.5pt} \diagup \hspace{-10pt}  \diagdown}

 \def\del{\partial}
 
 \def\bt{\bullet}
 \def\3dt{\dot{3}}
 
 \let\la=\label

 {}

 \def\bd{\begin{document}}
 \def\ed{\end{document}}
 \def\bea{\begin{eqnarray}}\def\barr{\begin{array}}\def\earr{\end{array}}
 \def\eea{\end{eqnarray}}
 \def\ft#1#2{{\textstyle{{\scriptstyle #1}\over {\scriptstyle #2}}}}
 \def\fft#1#2{{#1 \over #2}}
 \newcommand{\eq}[1]{(\ref{#1})}
 \def\eqss#1#2{(\ref{#1}-\ref{#2})}
 \def\eqs#1#2{(\ref{#1},\ref{#2})}
 \def\det{{\rm det\,}}
 \def\tr{{\rm tr}}\def\Tr{{\rm Tr}}
\def\ln{{\rm ln}}
\def\Li{{\rm Li}}

\begin{document}

\thispagestyle{empty}

  \hfill{\today}

 \vspace{20pt}

 \begin{center}
 {\Large{\bf Aspects of  superconformal field theories in six dimensions}}
 \vspace{30pt}

 {P.J. Heslop}\\[20pt]

 {\sl Institut f\"ur Theoretische Physik
der Universit\"at Leipzig}\\

 \vspace{60pt}

 \end{center}

 {\bf Abstract}

We introduce the analytic superspace formalism for six-dimensional
$(N,0)$ superconformal field theories. Concentrating on the $(2,0)$
theory we write down the Ward identities for correlation functions in
the theory and show how to solve them. We then consider the four-point
function of four energy momentum multiplets in detail, explicitly 
solving the Ward identities in this case. We expand the four-point
function using both Schur
polynomials, which lead to a simple formula in terms of a single
function of two variables, and (a supersymmetric generalisation of) Jack
polynomials, which allow a conformal partial wave expansion. We then
perform a complete 
conformal partial wave analysis of both the free theory four-point
function and the AdS dual four-point function. We also discuss
certain operators at the threshold of the series a) unitary bound, and
prove that some such operators can not develop anomalous dimensions, by
finding selection rules for certain three-point functions.
For those operators which are not protected, we
find representations with which they may combine to become long.

 {\vfill\leftline{}\vfill \vskip  10pt 
 \baselineskip=15pt \pagebreak \setcounter{page}{1}

\section{Introduction}
Since the discovery of  the $AdS/CFT$
correspondence~\cite{Maldacena:1998re,Witten:1998qj,Gubser:1998bc}
relating conformal field theories to supergravity, string theories or
M theories on an $AdS\xz S$ background, there
has been rapid progress in the investigation of conformal field
theories in dimensions larger than two.

The bulk of the analysis on the superconformal field  theory side of
the corespondence has concentrated on the
four dimensional $N=4$ supersymmetric Yang-Mills theory (SYM) and in
particular its correlation functions
(see~\cite{Aharony:1999ti,D'Hoker:2002aw,adsscft}
 for reviews).
This theory is of great interest for a number of reasons. It
has the largest possible amount of flat space supersymmetry in four
dimesnions, it is 
uniquely determined by the coupling constant and the gauge group and
it is superconformally invariant even as a quantum
theory~\cite{Sohnius:1981sn,Brink:1983pd,Howe:1984sr2}. In
particular, it is
conjectured to be dual to
IIB string theory on $AdS_5 \xz S^5$ and so both sides  of the AdS/CFT
conjecture are at least in principle well-defined theories, allowing
the possibility of testing the conjecture.

Less well understood is  the conjectured duality between M theory on
$AdS_7 \xz S^4$ and a six dimensional superconformal field theory with
$(2,0)$ supersymmetry. Neither side of this  conjecture is well
understood. On the  $AdS$ side  one knows little about M theory beyond
its low energy limit, supergravity. On the field theory side one knows
little beyond the free theory, that of the $(2,0)$ tensor
supermultiplet, which was first written down in~\cite{tensor} and
reformulated in a suitable harmonic superspace
in~\cite{onharm,howe6d}. This is believed, however to be the world
volume theory of the M theory 5-brane at low energies. 
Superconformal symmetry however provides a 
possible way
in to the study of the six dimensional conformal field  theory, which
can be compared with supergravity results and indeed
previous work in this area can be found
in~\cite{Park:1998nr,Corrado:1999pi,Eden:2001wg,AS}. 

Harmonic and analytic superspaces in four space-time dimensions were
introduced in~\cite{gikos}. A superfield on analytic superspace which
is Grassmann analytic and analytic on the internal variables can also
be thought of as an unconstrained superfield (but still analytic in the
internal coordinates) on analytic superspace which has a
reduced number of odd coordinates. This is similar to the way in which
a chiral superfield can be written as an unconstrained superfield on
chiral superspace. The general theory of such superspaces realised
as coset spaces of complexified superconformal groups was
developed in \cite{hartwell1,hartwell2} (see also
\cite{Luke}).

The analytic  
superspace formalism is particularly well suited for the study of
superconformal symmetry in four dimensions.
The advantages of using analytic superspaces  are firstly that the full
superconformal symmetry is manifest~\cite{Galperin:1985zv,
  hartwell1,hartwell2}.
Secondly, if one takes analytic superspace seriously as a (complex)
superspace (rather than
simply considering Grassmann analytic superfields on harmonic
superspace) one can give all superfields on analytic superspace (even
long ones). In general the superfields will transform linearly under
finite 
dimensional irreducible representations  of supergroups (so they carry
superindices) and remarkably
one finds that all irreducible unitary 
superconformal representations can be given in this way and that
furthermore they are all unconstrained (other than than the
requirement of analyticity
in all the complex variables)~\cite{paris,sindices}.
Furthermore (for even
number of supersymmetries $N$)
there is one 
analytic superspace which is singled out as  a (subset of the)
super Grassmanian of subspaces of half-dimension (that is of
dimension~$(2|N/2)$).
These are the natural analogues of
Minkowski space for the superconformal case. Indeed the Minkowski space
techniques of~\cite{os1,os2} can be adapted in the solution of Ward
identities.
 
In a series of
papers summarised in~\cite{adsscft} many aspects of $N=4$ SYM
involving half BPS operators were examined
using analytic superspace. Following the results
of~\cite{paris,sindices}
 these results were extended to more general operators on analytic
superspace~\cite{ops,3pt,u1y}. A summary of these latter results is
given in~\cite{dubna}. 

Of particular interest in the $N=4$ SYM case has been the four-point
function of energy-momentum multiplets~\cite{D'Hoker:1999jp,Eden:2000qp,Eden:2000bk,Arutyunov:2000ku,Arutyunov:2001im,Eden:2000mv,Eden:2000vb,Bianchi:2000hn,Arutyunov:2001mh,dos,4pt,asdo,Arutyunov:2003ad}. This is something which can be
calculated in the large N limit using gauged supergravity~\cite{D'Hoker:1999jp,Arutyunov:2000ku} yet it
contains within it information about all operators occurring in the OPE
of two energy-momentum operators (which for example includes operators dual to
string states.) It thus provides an important place
to both learn from and test the conjecture. Indeed the discovery of
new protected operators was found by this
method~\cite{Arutyunov:2000ku}. 
The  information
concerning operators in the OPE can be extracted from the four-point
function by a conformal partial wave  analysis (CPWA). This involves
decomposing the correlator into contributions from different operators
and was performed for the four-point correlator of energy-momentum
multiplets in $N=4$ SYM in~\cite{do,dos}. A CPWA was performed
for some higher charge half BPS operators in~\cite{asdo}.

In~\cite{4pt} four-point functions of energy momentum multiplets
and higher charge chiral primary operators were written down on analytic
superspace. Higher charge four-point functions were also written down
in~\cite{asdo} using a different method, and
were compared with results from IIB supergravity on $AdS_5\xz S^5$.

It is the purpose of this paper to apply analytic superspace
techniques to look at  the theory of
six-dimensional $(2,0)$ tensor supermultiplets. In~\cite{howe6d} this
theory was considered in the  harmonic superspace
description, half BPS 
operators were identified as representations and correlation functions
of half-BPS operators found. Using analytic
superspace (closely related to harmonic superspace) we will extend
 these results and show how
to find $n$-point correlation functions of any representations.

It is of particular interest to consider the four-point function of
energy-momentum multiplets in the six-dimensional theory and to
perform a CPW analysis on this. When considering the four-point
functions in $N=4$ SYM it proved useful to expand in terms of Schur
polynomials~\cite{4pt}. We will find that in the six-dimensional case
the Schur polynomial is also useful for finding a simple expression
for the correlator but one needs an expansion in a different basis
in order to relate this to the CPW. As for the Schur polynomials the
second basis  has an 
interpretation in terms of supergroup representation theory. 
Furthermore, in the
bosonic case this expansion reduces to the expansion in terms of Jack
polynomials used in six-dimensional CFT in~\cite{do6d} and so we can view
the supersymmetric expansion as the  
natural supersymmetric extension  of these Jack polynomials.

The full superconformal partial wave expansion can be found by lifting
from 
the bosonic case where the expansion is known~\cite{do6d}. Using this
we are able to perform a CPWA on both the free theory and  the large N
theory which is dual to gauged supergravity via  the AdS/CFT
correspondence. In particular we find the first $1/N^3$ corrections to
the dimensions of all operators in the OPE of two energy-momentum
multiplets. 

Finally we will consider operators in the theory which lie on the
threshold of 
the a)  unitary bounds  according to the classification of unitary
irreducible 
reps of~\cite{dobrev6d} (see section~\ref{sec:ubds}). In SYM operators
lying on the $N=4$ superconformal series a) bound come in two types:
those that develop 
anomalous dimensions and those that are protected. These protected
operators were first noticed by analysing the four-point function of
energy momentum tensors in the large N theory using AdS/CFT. This fact
was
later proved and generalised,
firstly by analysing the four-point function of
energy-momentum tensors~\cite{Arutyunov:2001mh}, and then by analysing
the Ward 
identities of three-point functions involving the  operator in
question and two half BPS 
operators~\cite{Arutyunov:2001qw}. 
The Ward identities give selection rules for the allowed
dimensions of the operator in question implying that some of them must
be protected.  

In~\cite{ops} another very simple 
proof was given which classifies an operator as protected or not based
purely on whether it is short or not (respectively) in the classical
interacting theory. In particular all operators which can be written
in terms of half BPS operators and saturate the unitary a) bound is
protected. Another recursive way to say this is that any operator
which can be 
written in terms of protected operators and which saturates the
unitary bounds is protected. Note that these operators need not lie in
the OPE of two 
half BPS operators. It is however, presumably possible to prove their
protectedness using the method of~\cite{Arutyunov:2001qw} by
  obtaining selection rules 
from  three-point functions of more
complicated operators than the half BPS ones. Indeed, as we shall see,
in the six-dimensional 
case it turns out that one is forced to take this approach.

In the six-dimensional case there are two complications. Firstly the
classical theory is not known and so the arguments of~\cite{ops} can 
not be applied. Secondly the operators which lie on the threshold of
the series a) bound do not lie in the OPE of two half BPS operators so
the selection rule arguments of~\cite{Arutyunov:2001qw} can also not be
applied straightforwardly~\cite{Eden:2001wg}. 
We therefore consider selection rules obtained from analysing the
three-point function of one half-BPS operator with one other protected
(but not half BPS) 
operator and a third operator. We again find the existence of
protected operators at the threshold of the unitary bounds.

The outline of the paper is as follows. In section~\ref{cor} we
introduce analytic superspace, consider the transformation of
operators, the superconformal Ward identities 
and we show how to solve these Ward identities. In section~\ref{4pt}
we examine the four-point function of energy momentum multiplets in
detail outlining the main results from the rest of the paper. In
section~\ref{bos} we consider expanding an invariant four point
function in Schur polynomials and Jack polynomials in a purely bosonic
CFT in six dimensions in order to illustrate the
techniques which we use in section~\ref{susy} for the full superconformal
case. In section~\ref{sec:cpwa} we perform a conformal partial wave
analysis  of the free four-point function and the AdS dual four-point
function. In section~\ref{crossing} we consider how crossing symmetry
acts on the four point function.
In section~\ref{rewritten} we rewrite the four-point function  in a
manner which allows more direct comparison with previous results and we
discuss the relation with previous results. Finally in
section~\ref{protected} we discuss operators lying at the threshold
of the unitary bound in the free theory. We find that some such
operators are protected and remain 
short and we find representations with which others may combine to
become long operators and hence develop anomalous dimensions. We
leave to the appendix technical details concerning the construction of
superspaces as supercosets of the superconformal group and how to find
the transformation of operators. We also discuss the general
construction of superconformal invariant $n$-point  functions  in the
theory in the appendix.

Whilst this manuscript was in preparation the preprint~\cite{dgs} appeared
which overlaps with the study of four-point functions
performed here. 

\section{Correlation functions}\la{cor}

The six-dimensional $(2,0)$ tensor supermultiplet consists of 5 scalar
fields transforming under the fundamental representation of the
internal group $SO(5)\sim USp(4)$, two Majorana-Weyl spinors in the
fundamental representation of $Usp(4)$, and a two form gauge field
with a self-dual 3-form field strength. In the free theory these
components can be 
packaged together on analytic superspace into a single analytic
superfield $W(X)$~\cite{howe6d}\footnote{Note here the close
analogy with the case of four-dimensional $N=4$ SYM where the
Yang-Mills multiplet can also be packaged into a single superfield on
analytic superspace.}. Analytic superspace has
half the number of odd coordinates as ordinary $(2,0)$ Minkowski
superspace, but also has an additional internal space $U(2)\bsh USp(4)$.
The local coordinates of analytic superspace combine into the
$(4|2)\xz(4|2)$ supermatrix 
\be
X^{AB}=\left( \ba{c|c} x^{\a \b} & \l^{\a b} \\
       \hline         -(\l^T)^{a \b}& y^{ab} \ea \right)\la{scoords1}
\ee
where the $x$s are the spacetime coordinates (in the spinor
representation), the $y$s are local
coordinates on the internal space and the $\l$s are the
odd coordinates. The indices $\a,\b$ are 4-component spinor indices,
$a,b$ are 2 component internal indices carrying the isotropy group
$U(2)$ of
the internal coset space. The matrix $X$ is generalised antisymmetric\footnote{ We choose $\a$ to be
an even index and $a$ to be an odd index so that for example
generalised anti-symmetrisation of $A$ and $B$ corresponds to
symmetrising $a$ and $b$ etc.} which
means that  $X^{AB}= -(-1)^{AB} X^{BA}$ and so 
\be
x^{\a \b} =-x^{\b \a} =0 \qquad y^{a b} = y^{b a}.
\ee
Here $x^{\a \b}=(\c_a)^{\a \b}x^a$ is the spinor representation of the
six-dimensional space-time coordinate $x^a$. Some properties of $\c$
matrices in six-dimensions are reviewed in appendix~\ref{sec:gamma}.

Using supercoset techniques (see section~\ref{has}) it is straightforward to show that an
infinitesimal superconformal transformation acts on these 
coordinates by
\be 
\d X= B + AX + XA^{sT} +XCX \la{dx}
\ee
where $A,B,C$ are all $(4|2)\xz (4|2)$ supermatrices, $B$ and $C$ are
generalised anti-symmetric, and $A^{sT}$ denotes the supertranspose of
the supermatrix $A$. The superfield $W$ transforms by
\be \d W=\cV W + \D W \la{Wtrans}\ee
where $\cV$ is the vector field generating the transformation, $\d
X^{AB}=(\cV X)^{AB}$ and
\be \D:=\str (A + XC).\ee

More generally operators will have superindices carrying an
irreducible representation of $\gg \gl(4|2)$, $\cR$ (see
section~\ref{sindices}) and they will transform as 
\be
 \d \cO^Q_{\cR}= \cV \cO^Q_{\cR} + \cR(A(X))
 \cO^Q_{\cR} + Q \D 
 \cO^Q_{\cR}\label{trans}
 \ee

where     $ A(X)= A +XC$.

 We can now write down and solve the Ward identities for
 correlators in the theory using similar techniques to those used for
 $N=4$ SYM in~\cite{3pt,u1y}.
The Ward identities for a general correlation function
\be <12\dots n>:=
<\cO^{Q_1}_{\unA_1}(X_1)\dots \cO^{Q_n}_{\unA_2}(X_n)> 
\ee
state that the correlator must be invariant under superconformal
transformations.
In other words
\be \d<12\dots n> = \sum_{i=1}^n (\cV_i + \cR_i(A_i)  + Q_i
\D_i)<12\dots n>=0 \la{ward} \ee
where $A_i:=A(X_i)\ D_i:=D(X_i)$. These can be easily solved as follows.

 For two-point functions we have the solution
 \be
 <\cT^{\unA} \cO^Q_{\unA}(1) \cO^Q_{\unB}(2) \cT^{\unB}> \ \propto \
 (g_{12})^Q
 \cT^{\unA} (X_{12}^{-1})^n_{\unA \unB} \cT^{\unB}
\ee

where $\cO^Q_{\unA}$ is an operator in the representation $\cR$
specified by a Young tableau with $n$ boxes which
is carried by the multi-index $\unA$, $\cT^{\unA}$ is an arbitrary
tensor also carrying the representation $\cR$ and we define
\be
(X_{12}^{n})_{\unA \unB}:=(X_{12})_{A_1B_1}\dots (X_{12})_{A_nB_n}.
\ee
 We have also introduced here the propogator $g_{12}$ which is
the two point function of $W$s in the free theory:
 \be <W(1)W(2)>\propto g_{12}=\sdet (X_{12}^{-1})={\hat y_{12}^2\over
x_{12}^4} \ee
where $X_{ij}:=X_i-X_j$ and where $\hat y=y+\l^T x^{-1} \l$.

The formula for the three-point function is as follows
\be \la{g3pt} \ba{rcl}
 \cT^{\unA_1}\cT^{\unA_2}\cT^{\unA_3}< \cO^{Q_1}_{\unA_1} \cO^{Q_2}_{\unA_2} \cO^{Q_3}_{\unA_3}>&
 \propto&(g_{12})^{Q_{12}}(g_{23})^{Q_{23}}
 (g_{31})^{Q_{31}} \cT^{\unA_1}\cT^{\unA_2}\cT^{\unA_3} \qquad \xz \\
 \\ && (X_{12}^{-1})^{n_2}_{\unA_2 \unB_2}
 (X_{13}^{-1})^{n_3}_{\unA_3 \unB_3} \xz
\ t(X_{123})^{\unB_2 \unB_3}_{\unA_1}
 \ea \ee
where $X_{123}=X_{12}X_{23}^{-1} X_{31}$, $Q_{ij}:=\half(Q_i+Q_j-Q_k);
k\neq i,j$ and where $t(X_{123})$ is a monomial of $X_{123}$ and its
inverse with the index structure as indicated. Note that in general
$t$ is not unique and one must take a linear combination of terms
(see section~\ref{protected} for an example of this.)

Similar formulae can be found for the higher point functions and
the general formula can be written schematically as~\footnote{see~\cite{u1y}
for the analogous formula in $N=4$ SYM}  
\be
<12..n> =P\Pi_{j=2}^n \cR_j(X_{1j}^{-1})\sum_t
t^{\cR_2\ldots \cR_n;\cR'_2\ldots \cR'_n}_{\cR_1\cR'_1}  F_t.
\ee
Here the sum is over all possible tensors $t$, $P$ denotes an
appropriate propogator factor, while $\cR_i$ are the representations
of the $\gg \gl(4|2)$ algebra under which the operators transform. Each
$F_t$ is an arbitrary function of superconformal invariants which we
discuss in appendix~\ref{invariants}. The correlator is further
restricted by demanding analyticity in all internal variables.
In the next section we discuss in detail the the four-point function of
four energy-momentum multiplets.

\section{Four-point functions}\la{4pt}

In~\cite{howe6d} $W$ was used to
generate a family of superfields in the free theory $A_p:=W^p$ which
were shown to be
superconformal. We wish to consider four-point functions of the
supercurrent $T:=A_2=W^2$
\be <T(X_1)T(X_2)T(X_3)T(X_4)>.\ee 
Although we do not know the interacting
theory we can still examine the Ward identities for correlation
functions in the theory.

The superconformal Ward identities are
\be \sum_{i=1}^4 \left( \cV_i + 2 \D_i\right) <TTTT>=0. \ee

The four-point function of four energy-momentum multiplets solving
these Ward identities can be
written on analytic superspace as
\be
<TTTT>=(g_{12}g_{34})^2 \xz \cI \la{4PT}
\ee
where $\cI(X_1,X_2,X_3,X_4)$ is invariant
under superconformal transformations. The
invariant is further restricted by insisting that it is analytic in
the internal coordinates. This is because $T$ is a polynomial in $y$
and hence analytic in $y$ so the right-hand side of~\eq{4PT} must be also.

We consider the problem of finding
such a function $\cI$. The analysis begins  similarly to
the four dimensional case which can be found in~\cite{adsscft,u1y}. Translation
invariance $\d X_i= B$ 
requires the function to depend only on difference variables
$X_{ij}$. Invariance under $C$ 
 requires the function to
depend only on 
the differences $X^{-1}_{ij}-X^{-1}_{ik}$ ie only on the variables
$X_{ijk}:=X_{ij} X^{-1}_{jk} X_{ki}$ which transforms as  
\be \d X_{ijk}=A_i X_{ijk} + X_{ijk} A^{sT}_i \ee
where $A_i:=A+X_iC$ (the variables $X_{ijk}$ are in fact the
negative inverses of the differences $X^{-1}_{ij}-X^{-1}_{ik}$). For
the four-point function 
there are only two independent variables $X_{ijk}$ which we choose to
be $X_{213}$ and $X_{243}$. Now change variables so that
our four-point function depends on the two variables $X_{213}$ and 
\be
Z:=-X_{213}X_{243}^{-1}=X_{21} X^{-1}_{13} X_{34} X_{42}^{-1}.
\ee
These  variables transform as
\bea \d X_{213}&=&A_2 X_{213} + X_{213} A^{sT}_2\la{dx213}\\
     \d Z&=&A_2 Z - Z A_2.
\eea
Note that $Z$ transforms in the adjoint representation of $\gg\gl(4|2)$.
We use this residual symmetry~\eq{dx213} to set the variable $X_{123}$ to the
following form
\be
X_{213}=K:=
\left( \ba{cc|c} 0&1&0\\-1&0&0\\ \hline 0&0&1 \ea \right)\la{C}
\ee
and then the remaining symmetry preserving $K$ is 
\be
\go\gs\gp(2|4)=\{A_2:A_2 K +K A^{sT}_2=0\}.\la{osp}
\ee

So we have reduced the problem of finding a four-point function
invariant under the conformal group to one of finding a function 
of $Z$ invariant under
the adjoint of $\go\gs\gp(2|4)$. The finite version of this 
 is invariance under the adjoint action of  the group $OSp(2|4)$,
\be Z\mapsto G^{-1} Z G\ee
where $G\in OSp(2|4)$.

There are now two bases which prove useful in expanding the
invariant function $\cI$. 
Firstly we use the basis provided by $GL(4|2)$
Schur polynomials.  
Schur polynomials (also known as characters) are simply (super)traces
of representations $S_{\cR}(Z)=\str(\cR(Z))$ where $\cR$ denotes a
  finite dimensional irreducible representation of $GL(4|2)$ (and
  hence a finite dimensional representation of  $OSp(2|4)$). The
allowed  representations $\cR$ for $<TTTT>$ (that is representations which
give a correlation function which is analytic in all the internal
variables) 
can be
described by a $GL(4|2)$ Young tableau with only two rows.
We arrive at   the formula
\be
<TTTT>=(g_{12}g_{34})^2 \sum_{p,\cR} C_{p,\cR} \ (\sdet Z)^p
S_{\cR} (Z).\la{4schur}
\ee
Analyticity in the internal variables of the four-point function puts
restrictions on the allowed representations $\cR$ and the values of
$p$. This follows the similar argument in the four dimensional
case~\cite{4pt}. Indeed $\sdet Z$ has poles in $y_{12}$ and this gives
the restriction
$p\leq 2$. The term $S_{\cR}$ may contain poles in $y_{13}$ however
and in order to remove these we require that $p\geq r$ where $r$ is
the number of rows of the Young tableau of $\cR$. So we find that
for representations with two rows we must have $p=2$, for those with 1
row we can have $p=1,2$ and for the trivial representation we may have
$p=0,1,2$.

Using this basis we can write the correlator in an explicit form as
follows. We use the remaining $OSp(2|4)$ symmetry $G$ to transform $Z$
and bring it into the 
diagonal form
\be \la{diagz}
Z={\rm diag}(X_1,X_2,X_1,X_2 | Y_1,Y_2).
\ee
So we can write the invariant four-point function in terms of the
eigenvalues $X_1,X_2,Y_1,Y_2$. Note that the the fact that the $X_i$
eigenvalues repeat simply comes from the fact that $Z$ wqas
constructed using antisymmetric supermatrices. We will give explicit
formulae for the Schur polynomials in terms of 
these variables below, but for now we simply state that
they can be used to show that the entire correlator can be written in
terms of a single function of two variables in the form
\be
<TTTT>={(g_{13}g_{24})^2\over (X_1-X_2)^4}\left({\bf \D} \left(\cS F(X_1,X_2)\right)+ \cS_1^2
F(X_1,X_1)
+\cS_2^2 F(X_2,X_2)\right)\la{4ptcls}
\ee
where $\cS:=(X_1-Y_1)(X_1-Y_2)(X_2-Y_1)(X_2-Y_2)$ and
$\cS_i=(X_i-Y_1)(X_i-Y_2)$ and ${\bf \D}$ is defined in~\eq{del}}.

There is another basis for the invariant function, however,  which allows us to
more easily relate the four-point 
function to the operators 
appearing in the OPE of two energy momentum multiplets and hence to
perform a conformal partial wave analysis. This second basis is given by
\be
T_\cR(Z):=\cY_{\cR}(K^n)_{\unA}(W^n)^{\unA}\la{TR}
\ee
where $\unA$ is a multi index put into the representation $\cR$ (which
is given by a Young tableau with $2n$ boxes)
using a  suitably normalised Young operator $\cY_{\cR}$. The Young
operator is defined similarly to the standard purely bosonic case the
only difference being that (anti-)symmetrisation is generalised. The
normalisation will be defined implicitly later when we give explicit
formulae for $T_{\cR}$ (for example in~\eq{tmn}). Here $K$ is defined
in~\eq{C} and $W$ is defined by $Z=-K W$.
In the purely bosonic case (the supersymmetric case can be
straightforwardly reduced t the bosonic case simply by taking all
superindices to be 
usual six-dimensional Weyl spinor indices, and hence ignoring the
internal indices) the basis 
elements $T_\cR(Z)$ reduce to the Jack polynomials (up 
to a factor) which were used in six
dimensional conformal field theory in~\cite{do6d}. The allowed
representations $\cR$ are specified by Young tableaux with four rows
with the first two rows of equal length and the second two rows of
equal length. For this expansion one obtains a similar formula
to~\eq{4schur} 
\be
<TTTT>=(g_{12}g_{34})^2 \sum_{p,\cR} C_{p,\cR} \ (\sdet Z)^p
T_{\cR} (Z).\la{4ptT}
\ee
with $p\leq 2$ and  $p\geq r/2$ where $r$ is
the number of rows of the Young tableau of $\cR$. So we find that
for representations with four rows we must have $p=2$, for those with 2
rows we can have $p=1,2$ and for the trivial representation we may have
$p=0,1,2$.

This latter formula facilitates a conformal partial wave analysis of the
four-point function. The OPE for two energy-momentum tensors $T$ is given
by
\be T(1)T(2) = \sum_{\cR_{MN}} {A_{TT\cO}\over (-8)^{N}C_{\cO\cO}}\,(g_{12})^{2-{q\over2}}
(X_{12}^{M+N})^{\unA}\, \cO^{q}_{\unA}(2) + \ldots \la{sOPE} 
\ee
The dots denote contributions of descendants of the primary fields
$\cO^q_{\unA}$, $\unA$ is a multi-index containing $2M+2N$ indices and
$(X_{12}^{2M+2N})^{\unA}:=X_{12}^{A_1A_2}\dots
X_{12}^{A_{2N+2M-1}A_{2N+2M}}$. The operator  
$\cO^{q}_{\unA}$ carries  the
tensor representation $\cR$ by having $2N+2M$ indices symmetrised
according to the Young tableau (with $2N+2M$
boxes, $M$ boxes in each of the first two rows and $N$ in each of
the third and fourth rows see~\eq{yt}) corresponding to
$\cR$. $C_{\cO\cO}$ is the coefficient for the two point function of
two operators $\cO^q_{\unA}$ defined in~\eq{s2pnt}, $A_{TT\cO}$ is the
coefficient of the three-point function defined in~\eq{3pt} whilst 
the numerical 
factor is present  in order to
reconcile the above definition of the OPE with the ordinary one in the
bosonic case see section~\ref{sec:pw}.  
Each primary field $\cO^q_{\unA}$ in this expansion carries charge
$q=d/2-j/2$ where $d$ is the
dilation weight and $j$ is the spin quantum number 
of the superconformal representation under which the operator
transforms. 
As seen in the previous section, in general an operator on analytic
superspace will also be 
a tensor (or quasi-tensor) field (indicated by the multi-index $\unA$)
carrying $2n$ superindices and will
transform under finite-dimensional irreducible representations of
the $GL(4|2)$ group which act on the superindices $A,B,\dots$. All of
the indices must be covariant (subscript) in order to be unitary and
the allowed representations must be those available in the
decomposition of  $
(X_{12}^n)^{\unA}$ into irreducible representations. Since the
building blocks, $X_{12}^{AB}$, 
for the Young tableaux are  antisymmetric
(corresponding to a Young tableau with 1 column and 2 rows) this
means the allowed  Young tableaux must have their first two rows of equal
length, and their second two rows of equal length (analysis of the
dependence on the internal coordinates shows that the Young tableaux
can have no more than $q$ rows.)

An important formula is that contribution of an operator
$\cO^{q}_{\cR}$ to the 
four-point function has the form
\be <TTTT> \sim {{(A_{TT\cO})^2 \over C_{\cO\cO}}} (g_{12} g_{34})^2
(\sdet Z)^{q/2}\sum_{\cR'}  C_{\cR'} T_{\cR'} (Z)
\la{scpw}\ee
where we sum over all representations $\cR'$ which have a Young
tableau with a valid form that
contains the Young tableau of $\cR$ (ie the Young tableau of $\cR'$
can be obtained by adding boxes to that of $\cR$.)
Here $A_{TT\cO}$ is the
coefficient of the 3-point function $<TT\cO>$
\be
<TT\cO^q \cdot \cT>=(-3)^N
A_{TT\cO}\,g_{12}^{2-q/2}g_{13}^{q/2}g_{23}^{q/2}\,
(X^{-1}_{312})_{\unA}\cT^{\unA}\la{3pt}
\ee
where $X_{312}=X_{31}X^{-1}_{12}X_{23}$,
$C_{\cO\cO}$ is the coefficient of the two point function of two operators $\cO^q_{MN}$
\be \la{s2pnt}
< \cO^Q_{\unA}(1) \cO^Q_{\unB}(2) \cT^{\unB}> \ =(4!)^N C_{\cO\cO} \
 (g_{12})^Q (X_{12}^{-n})_{\unA \unB} \cT^{\unB}
\ee 
and $\cT^{\unB}$ is an arbitrary
tensor carrying the same representation $\cR_{MN}$ carried by the operator
$\cO^q_{\unA}$. Indeed the first term in the expansion~\eq{scpw} can
easily be verified.  One performs an OPE~\eq{sOPE} on the four-point
function at points $X_3,X_4$ and
keeps only leading order terms in $X_{34}$ and only contributions from
the operator $\cO^q_{\cR}$ to  obtain
\bea
<TTTT>&\sim& {A_{TT\cO}\over C_{\cO\cO}} (g_{34})^{2-q/2}
\cY_{\cR}(X^{M+N}_{34})^{\unA}<TT\cO^q_{\unA}> +\dots\\
&\propto&{(A_{TT\cO})^2\over C_{\cO\cO}}(g_{12}g_{34})^2 (\sdet
Z)^{q/2}\cY_{\cR}(X^{M+N}_{34})^{\unA}(X^{-1}_{312})_{\unA}+\dots\\
&\propto&{(A_{TT\cO})^2\over C_{\cO\cO}}(g_{12}g_{34})^2 (\sdet
Z)^{q/2}T_{\cR}(Z)+\dots\eea 
where in the second line we have used~\eq{3pt} and in the third line
we have used the definition~\eq{TR} and the fact that
$T_{\cR}(X_{321}^{-1}X_{34})=\cT_{\cR}(Z)+\dots$ where the
dots represent terms of higher order in $X_{34}$.

We have thus obtained the first term in the expansion of~\eq{4ptT} and
hence motivated the appearance of the object $T_{\cR}$ in relation to
a conformal partial wave expansion.
In fact as we
shall see the numerical coefficients $C_{\cR}$ in~\eq{4ptT} can be
found using the results of~\cite{do6d}.

Furthermore, by relating the Schur polynomials to $T_{\cR}$ and using the relation
to the OPE~\eq{scpw} one finds that the
single two-variable function $F(X_1,X_2)$ splits as follows:
\be
F(X_1,X_2)=\l\ 
G(X_1,X_2)+{f(X_1)-f(X_2)\over \l}+{g(X_1)-g(X_2)\over
  X_1X_2\l}+{A\over X_1^2X_2^2}
\ee
where $\l:=X_1-X_2$. Here $A$ is the contribution of the identity
operator in the OPE, $g(X)$ 
gives  the contributions of operators with $q=2$ (all are short),
$f(X)$ has contributions from short operators with $q=4$ only and
$G(X_1,X_2)$ has contributions from all operators with $q=4$ (short
and long)\footnote{We define a long operator to be one which
  can be given as a superfield on Minkowski superspace which has a full expansion in
  odd-coordinates with independent coefficients. A short operator is
  an operator which is not long.}.

In summary, the four-point function of four energy-momentum multiplets
can be written on analytic superspace in the simple closed form
of~\eq{4ptcls} in terms of a single function of two variables. The
contributions of different types of operators in the OPE of two $T$'s
can be isolated and using the results of~\cite{do6d} a complete CPWA can
be performed. Indeed the conformal partial wave analysis is performed
explicitly in section~\ref{freecpw} for the free theory and in
section~\ref{lncpw} for the AdS dual theory.

\section{Purely bosonic case}\la{bos}

We will mainly be interested in the six-dimensional $(2,0)$
supersymmetric theory, but the formalism can be applied
straightforwardly to the case of $(n,0)$ supersymmetry for any $n$
including the purely bosonic case $n=0$. Therefore we firstly use this
simpler case to
illustrate the techniques. In the bosonic case the superindex $A$ becomes
a 6d spinor index $\a=(1,2,3,4)$, and the supercoordinates $X^{AB}$
of~\eq{scoords1} become the usual six-dimensional coordinates
$x^{\a\b}$ in the spinor representation corresponding to the upper
left block of $X$ in~\eq{scoords1}. The conformal group acts as
in~\eq{trans} if we regard all supermatrices to be
4$\xz$4 matrices corresponding to the upper left blocks of their
supermatrix counterparts. A scalar field of dimension 1
transforms like $W$ in~\eq{Wtrans} and we consider the four-point
function of fields with dimension $2$ which we denote
$T$ by analogy with the supersymmetric case. 
The four-point function reads $<TTTT>=(g_{12}
g_{34})^2 \xz \cI$ with $\cI$ invariant under conformal
transformations. We then consider the problem of finding $\cI$ and
following the arguments of equations~(\ref{4PT}-\ref{4ptT}) we arrive at
a function of the matrix $Z={\rm diag}(X_1,X_2,X_1,X_2)$ and   
two alternative expansions, the $T$ expansion or the $S$ Schur
polynomial expansion which we now proceed to find explicit
expressions for.

We wish to define the basis $T_{\cR}(Z)$ in the bosonic case
($Z=\rm{diag}(X_1,X_2,X_1,X_2)$ in this case.)
We firstly consider the simplest 
representation which is the antisymmetric representation $\cR=${\tiny$
\yng(1,1)$}. The basis element corresponding to this representation is
simply 
\be
T_\cR(Z)=C_{\a \b}W^{\a \b}=\tr(Z)=2X_1+2X_2.
\ee
One can also find the basis
elements corresponding to Young tableaux with two rows of equal length
$m$

\be
\cR_m:=\overbrace{\yng(7,7) }^m.
\ee

This is given by   

\be
T_{\cR_m}(Z)=C_{(\a_1 |\b_1|} ...C_{\a_m) |\b_m|}  W^{\a_1
  \b_1 }  \dots W^{\a_m \b_m }=\tr(\cR'_m(Z))\la{symbos}
\ee

where we symmetrise over the $\a_i$ indices but not the $\b_i$ indices.
Here $\cR'_m$ is the completely symmetric representation with $m$
boxes
and $\tr(\cR'(Z))$ is simply the ordinary $GL(4)$ Schur polynomial of
$Z$. 
An explicit formula for these in terms of the eigenvalues $X_1,X_2$
can be found and is
given by the formula
\be
t_{m}(Z):=(X_1-X_2)^3 \ T_{\cR_m}(Z)=
(m+1)(X_1^{m+3}-X_2^{m+3})-(m+3)X_1X_2(X_1^{m+1}-
  X_2^{m+1})     \la{ta}
\ee

Although these are a priori only defined for $m\geq0$ we will allow
$m$ to take any integer value. We may then note the following
important special cases

\be
t_{0}(Z)=1\la{t0} \qquad
t_{-1}(Z)=
t_{-2}(Z)=
t_{-3}(Z)=0\la{t-3}
\ee

and also the relation

\be
t_{-m}(Z)=-(X_1X_2)^{2-m}t_{m-4}.\la{relation}
\ee

An explicit formula for the most general representation (occurring in the OPE of two
scalars) is given by
\bea &
\cR_{mn}=\underbrace{\yng(0,0,4,4)}_{n}
\hspace{-51.2pt}\overbrace{\yng(8,8,0,0)}^{m}&   \la{yt}
\\[10pt] &\Downarrow& \nonumber \\[10pt]
&\ba{rcl}t_{mn}(Z)&:=&T_{\cR_{mn}}(X_1-X_2)^3=(X_1 X_2)^n t_{m-n}(Z) \\
                 &=&(m-n+1)X_1^{m+3}X_2^n-(m-n+3)X_1^{m+2}X_2^{n+1} - (X_1 \leftrightarrow X_2)
    \ea 
&\la{tmn}\eea
Notice that the $T_{\cR_m}$ are precisely (up to an overall factor)
the Jack polynomials used in~\cite{do6d} in this context.

An invariant four-point function can be expanded in the basis $t_{mn}$.
The relation~\eq{relation} implies that 
\be
t_{m-2\ n}=-t_{n-2\ m}
\ee
and so although 
a priori $t_{m-2\ n}$ is only valid for $m-2 \geq n \geq 0$ (for the Young tableau~\eq{yt}) to have the correct shape) 
we will extend this to any values in the range $m,n\geq0$ by noting this
symmetry (and also the fact that
$t_{n-1,n}=t_{n-2,n}=t_{n-3,n}=0$.)

\subsection{Partial wave expansion}\la{sec:pw}

In~\cite{do6d} the exact
expression for the contribution
of an operator in the OPE of two scalars to the four-point function of
four such scalars in a purely bosonic six-dimensional conformal field
theory was found as an expansion in Jack polynomials. This expression is
known as the conformal partial wave and we 
briefly review this here for later use in the supersymmetric case. The
formulae for the OPE and three-point functions given in~(\ref{sOPE},\ref{3pt}) can 
be applied straightforwardly to the purely bosonic case by simply
letting all superindices become spinor indices and the supermatrix $X$
become the space-time variable $x$ in spinor notation. 
The energy-momentum tensor $T$ becomes a scalar field $\phi(x)$ with
dilation weight 2 and an operator $\cO^q_{M,N}$ becomes isomorphic to
an operator $\co_{M+q,N+q}$. We denote a general operator in the
bosonic theory with dilation weight $M+N$ and spin $M-N$ by $\co_{MN}$.
From now on in this section we set $q=0$ without loss of generality in the
bosonic case. The isomorphism is given explicitly as
\be
(\cO^0_{MN})_{\ua}= \e_{\a_1\a_2\a_3\a_4}\dots \e_{\a_{4\!N\!-\!3}\dots\a_{4\!N}}
({ \co}_{MN})_{\a_{4\!N\!+\!1}\dots \a_{2\!M\!+\!2\!N}}.
\ee
Here $\cO^0_{MN}$ carries the tensor representation indicated by the
Young tableau~\eq{yt} and is therefore completely antisymmetric on the
$N$ columns of the Young tableau which are therefore proportional to
$\e$ tensors as indicated. The operator $\co_{MN}$ has $2(M-N)$ spinor
indices and is simply the spinor representation of a field with  $M-N$
symmetric traceless space-time indices.
The OPE~\eq{sOPE} then reads
\bea
\phi(x_1)\phi(x_2)&\sim& {(-8)^{-N}A_{\phi\phi\cO}C^{-1}_{\cO\cO}}
(x_{12}^2)^{-4}\,
(x_{12}^{M+N})^{\ua}(\cO^0_{MN})_{\ua}(2)+\dots.\\
&=&A_{\phi\phi\cO}C^{-1}_{\cO\cO}
(x_{12}^2)^{N-4}\,
(x_{12}^{M-N})^{\ua}(\co_{M N})_{\ua}(2)+\dots.\la{OPE2}
\eea
This explains the presence of the factor $(-8)^N$ in~\eq{sOPE} which
is cancelled in~\eq{OPE2} by the factor coming from the $N$
applications of~\eq{x^2}.

If we similarly take the expression for the three point
function~\eq{3pt} and define $\ct^{\ua}$, isomorphic to $\cT^{\ua}$, by
\be
\cT^{\ua}=\e^{\a_1\a_2\a_3\a_4}\dots
\e^{\a_{4\!N\!-\!3}\dots\a_{4\!N}}
\ct^{\a_{4\!N\!+\!1}\dots\a_{2\!N\!+\!2\!M}} 
\ee
then $\cO^0_{MN}\cdot\cT=(4!)^N{\co}_{MN}\cdot \ct$ and
\bea
<\phi\,\phi \,\co_{MN}\cdot\ct>&=&\left(-{8}\right)^{-N} A_{\phi\phi\cO} \,(x_{12}^{2})^{-4}\, 
(x^{-1}_{312})^{M+N}_{\ua}\cT^{\ua}\\
&=& 
A_{\phi\phi\cO} \,(x^2_{12})^{N-4}(x_{23}^{2})^{-N}(x_{13}^{2})^{-N}\, 
(x^{-1}_{312})^{M-N}_{\ua}\ct^{\ua}.\la{3pt2}
\eea
Again we see that the numerical factor defined in~\eq{3pt} cancels. 

We now consider the conformal partial wave expansion of the four-point
function $<\phi \phi \phi \phi>$. Firstly consider this in
the limit $x_3\rightarrow x_4$. To leading order in $x_{34}$ the
contribution of the operator $\co_{MN}$ to this four-point function
can be found by performing the 
OPE on $\phi(3)\phi(4)$ using~\eq{OPE2} and then using~\eq{3pt2}. We
obtain that the conformal partial wave expansion for the operator
$\co_{MN}$ has the form
\bea
&& \!\!\!\!{A_{\phi\phi\cO}\over C_{\cO\cO}}<\phi \phi (\co_{M N})_{\ua}>
 (x_{34}^{M\!-\!N})^{\ua} (x_{34}^2)^{N-4}+ \dots\\ 
&=& \!\!\!\!{(A_{\phi\phi\cO})^2\over C_{\cO\cO}}{1\over
    (x^2_{12}x_{34}^2)^4}(z^2)^N
  \mbox{\small $(x_{312}^{-1})_{\a_1\b_1}\dots(x_{312}^{-1})_{\a_{M\!-\!N}\b_{M\!-\!N}}x_{34}^{(\ua_1|\ub_1|}x_{34}^{\ua_2|\ub_2|}\dots x_{34}^{\ua_{M\!-\!N})\ub_{M\!-\!N}}$}+\dots \\ 
&=& \!\!\!\!{(A_{\phi\phi\cO})^2\over C_{\cO\cO}}{1\over
    (x^2_{12}x_{34}^2)^4}T_{\cR_{MN}}(z) +\dots.\la{4pt3}
\eea
In the second line we have to symmetrise the indices on the $x_{34}$s
as indicated  and if we compare this with~\eq{symbos} we see the
appearance of the Jack polynomial $T_{\cR_{M-N}}$ which combines with
the $(z^2)^N$ to give $T_{\cR_{MN}}$ (see~\eq{tmn}.) In all four of
the equations above the dots indicate contributions from higher orders
in $x_{34}$.

The full conformal partial wave (including all orders in $x_{34}$)
corresponding to the operator 
$\co_{MN}$ was found in~\cite{do6d} and is given by
\be
<\phi(1) \phi(2)\phi(3)\phi(4)>\sim {(A_{\phi\phi\cO})^2\over
  C_{\cO\cO}}\  {1 \over (x_{12}^2)^{4}(x_{34}^2)^{4}} F_{MN}\la{cpwabos}
\ee
where $c$ is a constant and $F_{MN}$ is given by 
\be \ba{rcl}
F_{M\,N}&=&\sum_{m,n\geq0} c_{M,N}(m,n)T_{\cR_{M+m, N+n}}\la{FMN}\\ 
\\
c_{M,N+2}(m,n)&=&{M-N-1\over \m-1}\left(1-{2n\over (M-N-1)(\m+1)}-{2mn\over
  (M-N-1)(\m+1)(M+N)}\right){(M)_{m}^2 \over
(m)!\,(2M)_{m} } {(N)_n^2 \over n! \,(2N)_n}
\ea \ee
where $\m:=M+m-N-n$. Notice that here we normalise $F_{MN}$ so that
$c_{M,N}(0,0)=1$ and hence  
$F_{MN}=T_{\cR_{MN}}+\dots$ to be consistent with~\eq{4pt3}.

With this information, if we know the four-point
function we can work out the coefficient
$A^2_{\phi\phi\cO}/C_{\cO\cO}$ for all operators in the
OPE. Furthermore, from one loop four-point functions 
one can find the anomalous dimensions of operators (see
section~\ref{lncpw}). 

\subsection{Schur polynomials}

As well as expanding the invariant four-point function in the Jack
polynomials $T$ 
it is useful in the supersymmetric case to also consider an expansion
in terms of
Schur polynomials of
$Z$ which we denote $S_{\cR}(Z):=\tr(\cR(Z))$. This latter expansion allows
one to write the four point function in terms of a single two-variable
function in the supersymmetric case.   

In the bosonic case we know that~\eq{symbos}
\be
T_{\cR_m}=S_{\cR'_m}\la{T=S}
\ee
So we already know the Schur polynomials for single row Young
tableaux.

To find the
Schur polynomials of more complicated Young tableaux we use the
property that 
multiplication of Schur polynomials corresponds to the tensoring of the
corresponding representations. In other words
$s_{\cR}s_{\cS}=\sum_{\cT} d_{\cR \cS \cT} s_{\cT}$
where 
$d_{\cR \cS \cT}$ are the numbers in the decomposition of the
tensor product of $\cS$ and $\cR$ into irreducibles:
${\cR}\otimes{\cS}=\sum_{\cT}d_{\cR \cS \cT} {\cT}$.

Using the well-known rules for multiplying Young tableaux one can show
that $\cR'_{mn}=\cR'_{m}\cR'_n-\cR'_{m+1}\cR'_{n-1}$ where

\be
\cR'_{mn}=\underbrace{\yng(0,4)}_{n}
\hspace{-51.2pt}\overbrace{\yng(8,0)}^{m} 
\qquad \qquad \cR'_m=\overbrace{\yng(8)}^m \ee

and it follows that  

\be
S_{\cR'_{mn}}=S_{\cR_m}S_{\cR_n}-S_{\cR_{m+1}}S_{\cR_{n-1}}.\la{sab}
\ee

 Explicitly, using~(\ref{tmn},\ref{T=S}) one finds that

\be 
s_{m-1\, n}:=(X_1-X_2)^4\xz S_{\cR'_{m-1\, n}}=\ba{rl}
&\left(\left( m-n \right)\,X_1^{3 + m + n} - 
    \left( 2 + n \right) \,\left( 1 + m \right) \,X_1^{2 + m}\,X_2^{1 +
      n}  +\right.\\
&\left. \left( 1 + n \right) \,\left( 2 + m \right) \,X_1^{1 + m}\,X_2^{2 +
      n}\right) + X_1 \leftrightarrow X_2
\ea  \la{expl}\ee

which satisfies:
\be
s_{m-1\, n}=-s_{n-1\, m}.\la{santi}
\ee

So although $s_{m-1\, n}$ is a priori only defined for $m-1\geq n\geq
0$, one 
can extend it to the range $m,n \geq0$ using the above equation and
the fact that $s_{m\, m+1}=0$.

Equation~\eq{expl} can be rewritten:
\bea
s_{m-1\, n}&=&{\bf \D}  \ 
  \left({X_1^{m+2} X_2^{n+2}-X_1^{n+2} X_2^{m+2}\over X_1-X_2} \right)+
(m-n)(X_1^{3+m+n}+X_2^{3+m+n})
\la{s=f}\\[5pt]
{\bf \D}&:=&-(\del_1- \del_2 +\l\del_1\del_2)\l \la{del}
\eea
where $\l=X_1-X_2$.

One can expand a four-point function in the `s' basis as well as the
`t' basis and this can be expressed in terms of an antisymmetric
function of two variables $F$ as

\bea \la{sol}
\sum_{m,n\geq0}d_{mn}s_{m-1\,n}(X_1,X_2)&=&{\bf \D} F(X_1,X_2)+F(X_1,X_1)+F(X_2,X_2)\\
F(X_1,X_2)&:=&{1\over \l}\sum_{m,n\geq0}d_{mn}(X_1^{m+2} X_2^{n+2}-X_1^{n+2}
X_2^{m+2})
\eea
a formula which can be readily generalised to the supersymmetric case.

The `s' basis is related to the `t' basis by
\be
\l\,t_{m-2\, n}={(m-n+1)\,s_{m-2\,n}- (m-n-1)\,s_{m-1\ n-1} \over (m+1)(n+1) }. \la{ts}
\ee
This formula generalises directly to the supersymmetric case where it
is more useful than in the present context.

Before considering the supersymmetric case we would like to know to
what extent the function $F(X_1,X_2)$ uniquely defines the four-point
function. One can show that the general solution of 
${\bf \D}
F(X_1,X_2)+F(X_1,X_1)+F(X_2,X_2)=0,\ F(X_1,X_2)=F(X_2,X_1)$ is given by 
 \be \la{kernel2}
F(X_1,X_2)={f(X_1)-f(X_2)+X_1 X_2(g(X_1)-g(X_2))\over X_1-X_2}
\ee
with $f,g$ arbitrary functions.

\section{The supersymmetric case} \la{susy}

\subsection{Operators in the OPE of two energy momentum multiplets}

Before we consider the four-point function of four energy momentum
multiplets $T$ in the $(2,0)$ supersymmetric case,  we first classify
the operators which will appear in the OPE of two $T$s and hence in the
CPW expansion of the four-point function. 

The starting point for the classification of operators is the formula
for the OPE given in equation~\eq{sOPE} and the analysis sketched below that
equation shows that operators in the OPE must carry representations of
$GL(4|2)$ given by the Young tableaux:
\be \cR_{MN}=\underbrace{\yng(0,0,4,4)}_{N}
\hspace{-51.2pt}\overbrace{\yng(8,8,0,0)}^{M}.\la{YT}
\ee
Notice that these Young tableaux are the same as those defining
operators in the bosonic OPE (see~\eq{yt}) 
although of course the interpretation is different: in the bosonic
case one is considering $GL(4)$ representations whereas here we are
considering $GL(4|2)$ representations. However this point turns
out to be crucial when considering the CPW expansion since it enables
us to read off the supersymmetric CPW from the bosonic
one. In the supersymmetric case the operators also carry a charge
$q=0,2$ or $4$ and  the 
Young tableau can have no more than $q$ rows. 

The representations of $GL(4|2)$, $\cR_{MN}$ split into four classes:
$N\geq2$ are long representations (called typical in the mathematics
literature) and lead to long supermultiplets\footnote{In
fact the long  representations can have
non-integer $n$ a fact which allows long operators to develop
anomalous dimensions. For this one must use quasi-tensor
representations~\cite{3pt}. This will not concern us in the present
work.}; $N=1$ gives one class of short (or atypical)
representations; $N=0, M\neq0$ gives another class of short
representations; finally $M=N=0$ is the trivial representation. We
denote a general operator carrying charge $q$ and 
$GL(4|2)$ representation $\cR_{MN}$ by $\cO^q_{MN}$. We denote
component fields which carry a representation of the internal group
$USp(4)$ as well as Lorentz spin $M-M$ and dilation weight $M+M$ by
$\vf^{\rm \bf rep}_{M,N}$ where ${ \rm \bf rep}$ is the dimension of
the $USp(4)$ representation.  
There are seven classes of operators, given in table~\ref{table1}
along with the lowest component of the multiplet and the component
field obtained in taking the `bosonic limit' which we do in
section~\ref{sec:spwe} in order to find the superconformal partial
wave expansion.

\begin{table}[htbp]
  \centering
  $\ba{|lll|}\hline
\mbox{Superfield }  & \mbox{Lowest }  & \mbox{`Bosonic
limit' } \\[-5pt] 
& \mbox{component }&\mbox{component}\\
\hline \cO^4_{M, N\geq2} &\vf^{{\bf 1}}_{M+2,N+2}& \vf^{{\bf 55}}_{M+4,N+4}\\[5pt]
\cO^4_{M, 1} &\vf^{\bf 10}_{M+2,4}& \vf^{{\bf 55}}_{M+4,5}\\[5pt]
\cO^4_{M, 0} &\vf^{{\bf 14}}_{M+2,4}& \vf^{{\bf 55}}_{M+4,4}\\[5pt]
\cO^4_{0, 0} &\vf^{{\bf 55}}_{4,4}& \vf^{{\bf 55}}_{4,4}\\[5pt]
\cO^2_{M, 0} &\vf^{{\bf 1}}_{M,2}& \vf^{{\bf 14}}_{M+2,2}\\[5pt]
\cO^2_{0, 0} &\vf^{{\bf 14}}_{2,2}& \vf^{{\bf 14}}_{2,2}\\[5pt]
\cO^0_{0,0} &\vf^{{\bf 1}}_{0,0}& \vf^{{\bf 1}}_{0,0}\\\hline\ea$
  \caption{\small Operators in the OPE of two $T$s.}
  \label{table1}
\end{table}

\subsection{The  invariant four-point function in the supersymmetric
  case}\la{sexp}

We now consider the
four-point function of four energy-momentum multiplets in the $(2,0)$
supersymmetric theory.
Remarkably much of the formalism from the bosonic case goes through
fairly straightforwardly in the supersymmetric case also.

Firstly we wish to find formulae for the basis elements $T_{\cR}$ with
which we will expand the four-point function as in~\eq{4ptT}. The
simplest representation is the (generalised) anti-symmetric
representation $\cR=${\tiny $\yng(1,1)$}. The basis element corresponding
to this representation is
\be
T_\cR(Z)=K_{\unA \unB}W^{\unA \unB}=\tr(Z)=2X_1+2X_2-Y_1-Y_2.
\ee
As in the bosonic case, the basis elements $T_{\cR}$ of Young tableaux
with two rows of 
equal length $m$ coincide with the $GL(4|2)$ Schur polynomials of Z,
$\str(\cR'_m(Z))$, of representations with a single row (see 
eq~\eq{symbos}). 
These can be found explicitly by decomposing the supergroup $GL(4|2)$ into its
maximal bosonic subgroup $GL(4)\times GL(2)$.

The Schur polynomials respect this decomposition in the following
sense. If the $GL(4|2)$ representation $\cR$ decomposes as
$\cR\rightarrow \sum_{\cS,\cT}d_{\cR \cS \cT}\cS\otimes\cT$
under $GL(4|2)\supset GL(4)\times GL(2)$ (here $\cS$ is a $GL(4)$
representation 
and $\cT$ a $GL(2)$ representation and $\cS\otimes\cT$ is hence a representation of
$ GL(4)\times GL(2)\subset GL(4|2)$)
then the Schur polynomials satisfy
$s_\cR(Z)=\sum_{\cS,\cT}d_{\cR \cS \cT}s_\cS(X)s_\cT(Y)$
where 
\be
Z=\left(\ba{c|c}X&0\\ \hline 0&Y\ea\right).
\ee

Using this one can  write the supersymmetric Schur polynomials in terms of
the bosonic Schur polynomials. For example the Schur polynomials with
only one row in their Young tableaux are given by:
\be
S_m(Z)=s_m-s_{m-1}(Y_1+Y_2)+s_{m-2}Y_1 Y_2.
\ee
This formula is a priori only valid for $m\geq 2$ since $s_{m}$ is
only valid for $m\geq0$. However from~\eq{t-3} we see that the
formula  also gives the correct answer for $S_1$ and $S_0$ without
modification and also gives a vanishing $S_{-1}$
\be
S_0(Z)=1\qquad S_{-1}(Z)=0.\la{srelations}
\ee
Note that these relations distinguish the six dimensional case from the case of
four dimensional $N=4$ super Yang-Mills where Schur polynomials for
short representations had to be treated separately~\cite{4pt}. This in
turn leads to the requirement of extra functions of one variable in
the four-dimensional case whereas one only needs a single two variable
function in the six-dimensional case.

Similarly for Schur polynomials with two rows $S_{mn}:=\l^4
S_{\cR'_{mn}}$ decomposes naturally into its  
component purely bosonic Schur polynomials:
\be \ba{rcl}
S_
{m-1\, n}(X_1,X_2,Y_1,Y_2)&=&s_{m-1\,n}-(Y_1+Y_2)\,(s_{m-1\, n-1}+s_{m-2\,
  n})+\\&&Y_1Y_2\,(s_{m-1\,n-2}+s_{m-2\,n-1}+s_{m-3\,n})\,+
(Y_1^2+Y_1Y_2+Y_2^2)\,s_{m-2\,n-1}-\\&&Y_1Y_2(Y_1+Y_2)\,(s_{m-2\,n-2}+s_{m-3\,n-1})+Y_1^2Y_2^2\,s_{m-3\,n-2}
\ea \la{Smn}\ee
and using~\eq{santi} we find that 
\be S_{m-1\,n}=-S_{n-1\,m}\la{Santi}
\ee
and so we can extend the range of validity of $S_{mn}$ to $m\geq -1,\
n\geq0$.

Equation~\eq{Smn} can be rewritten (using~\eq{s=f}) in the form
\bea 
S_{m-1\,n}&=&{\bf \D} \left(\cS
  \xz\left({X_1^{m}X_2^{n}-X_1^{n}X_2^{m}\over X_1-X_2}\right)\right)\ba[t]{l}+
(m-n)(X_1-Y_1)^2 (X_1-Y_2)^2 X_1^{m+n-1}\\+(m-n)(X_2-Y_1)^2
(X_2-Y_2)^2X_2^{m+n-1}\ea\la{S=F}\\[10pt] 
\cS&:=&
\ba{l}X_1^2X_2^2-(Y_1+Y_2)X_1X_2(X_1+X_2)+Y_1Y_2(X_1^2+X_1X_2+X_2^2)+\\Y_1^2Y_2^2-(Y_1+Y_2)Y_1Y_2(X_1+X_2)+X_1X_2(Y_1^2+Y_1Y_2+Y_2^2).\ea\\
&=&(X_1-Y_1)(X_1-Y_2)(X_2-Y_1)(X_2-Y_2)\la{Snice}
\eea

It is also useful to note the two identities
\bea \la{Sm-1}
S_{m,-1}&=&{Y_1Y_2\over (X_1X_2)^2}S_{m+1\,0} \\
S_{-2,-2}&=&-S_{-3,-1}={(Y_1Y_2)^2\over (X_1X_2)^4}.\la{S-2}
\eea
which are crucial for allowing us to treat the contributions of  operators of
dimension 2 
and the identity operator in the same formula. 

Therefore an invariant four-point function $\cI$ expanded in the `S' basis
can be written in terms of a single function of two variables
$F(X_1,X_2)$ as 
\bea
\l^4\cI=\cF[F]:=\sum_{m,n \geq0}f_{m\,n}S_{m-1\, n}&=&{\bf \D} \left({\cS} \
  F(X_1,X_2)\right)+ \cS_1^2
F(X_1,X_1)
+\cS_2^2 F(X_2,X_2)\la{Sbas}\\
F(X_1,X_2)&=&\sum_{m,n \geq0} f_{mn}\,
{X_1^{m}X_2^{n}-X_2^{m}X_1^{n}\over X_1-X_2}
\eea
where $\cS_i=(X_i-Y_1)(X_i-Y_2)$.
This is the complete result for the four-point function of four
energy-momentum multiplets given in~\eq{4ptcls}.

In order to separate out correctly the contributions from short and long
operators to the four-point function, equation~\eq{scpw} tells us we
need to use the $T$ basis rather than the $S$ basis. In order to find
explicit expressions for the $T$ basis we note that it is related
to the Schur polynomial $S$ basis by  a similar formula to the bosonic
case (see~\eq{ts})
\be
\l\ T_{m-2\,n}= \left\{ \ba{ll} {(m-n+1)S_{m-2\,n}- (m-n-1)S_{m-1\ n-1} \over (m+1)(n+1) }
\qquad &m,n\geq 1\\ S_{m-2\ 0} &n=0
\ea \right.
\la{TS} 
\ee
In the supersymmetric case (unlike the bosonic case) $n=0$ has to be handled seperately
since our expression for $S_{mn}$~\eq{Smn} does not give 
$S_{m\ -1}=0$. From~\eq{Santi} we find that
this expression satisfies the symmetry
\be
T_{m-2\, n}=-T_{n-2\,m}\la{Tanti}
\ee
which can be used to extend the range of validity of~(\ref{TS}a) from
$m-2\geq n\geq 1$ to
the range $n\geq 1,\ m\geq 1$.

Equation~\eq{TS} can be rewritten, using~(\ref{S=F},\ref{tmn})
as
\be \ba{rcll}
\l\ T_{m-2\, n}&=& -{\bf \D}\, \left({S\over \l
    X_1X_2}{t_{m-2\ n}\over 
    (m+1)(n+1)}\right) \qquad &m,n\geq1\\[5pt]
\l\ T_{m-2\,0}&=&
{\bf \D}\left(S\ {X_1^{m-1}-X_2^{m-1}\over \l}\right)
\ba[t]{l}+(m-1)(X_1-Y_1)^2(X_1-Y_2)^2 X_1^{m-2}\\+(m-1)(X_2-Y_1)^2(X_2-Y_2)^2
X_2^{m-2}\ea \\[5pt]
&=&
-{\bf \D}\, \left({S\over \l X_1X_2}{t_{m-2\ 0}\over (m+1)}\right) +{m-1\over
  m+1}S_{m-1,-1} 
 \ea \ee

We also use the following which comes from~(\ref{S=F},\ref{Sm-1},\ref{S-2},\ref{TS}b)
\be \ba{rcl}
\l\,{Y_1Y_2\over X_1^2X_2^2}T_{m-2\,0}&=&S_{m-3,-1}\\
&=&{\bf \D}\left({S\over X_1X_2}
  {X_1^{m-1}-X_2^{m-1}\over \l}\right)\ba[t]{l}+(m-1)(X_1-Y_1)^2(X_1-Y_2)^2X_1^{m-4}\\
                                       +(m-1)(X_2-Y_1)^2(X_2-Y_2)^2X_2^{m-4}\ea
\ea
 \ee
The contributions of the last two terms of~\eq{S=F}
vanish for $m,n\geq1$. 

The  invariant function $\cI$ needed in the four-point function
expanded in the `T' basis consists of four terms all of which can be
written in the form~\eq{Sbas} in terms of a function of two variables in the
following way.
\be
\l\sum_{m,n\geq 1}d_{mn}T_{m-2\
  n}+\l\sum_{m\geq 0}d_{m}T_{m\,0}+\l {Y_1Y_2\over X_1^2X_2^2}
\sum_{m\geq 0}c_m T_{m\,0}+A{Y_1^2Y_2^2 \over
  X_1^4X_2^4}=\cF[F_1+F_2+F_3+F_4]\la{4ptTs}
\ee
where
\be
\ba{lcrcl}
F_1=\l G&\qquad&G(X_1,X_2)&:=&-{1\over \l^2 X_1X_2}\sum_{m,n\geq1}d_{m\,n}{t_{m-2\, n}\over
  (m+1)(n+1)}\\[5pt]
F_2=(f(X_1)-f(X_2)/\l&&
f(X)&:=&\sum_{m\geq0}d_{m}X^{m+1}\\[5pt]
F_3=(g(X_1)-g(X_2)/(X_1X_2\l)&&
g(X)&:=&\sum_{m\geq0}c_{m}X^{m+1}\\[5pt]
F_4=A/(X_1^2X_2^2)
\la{invform}
\ea
\ee

The split into $F_1,F_2,F_3,F_4$ is unique if we assume holomorphicity
of all functions. In this way, one can isolate  the contributions
of short operators in the OPE.

An important point to note is that in the supersymmetric case, by
writing the four-point function in the form $\cF[F]$ as 
in~\eq{Sbas}, the function $F$ is uniquely defined. Indeed
using~\eq{kernel2} one can show that the solution of $\cF[F]=0$ for
all values of $Y_1,Y_2$ is $F=0$.

\subsection{Superconformal partial wave expansion}\la{sec:spwe}

We are now in a position to give a complete superconformal partial wave expansion for
the four-point function of four energy-momentum multiplets. In order
to do this we consider the four-point function $<TTTT>$ in the limit
where $y_{12},y_{34}\rightarrow 0$ 
and where all analytic superspace odd coordinates vanish.
We call this the `bosonic limit'. In this limit 
the remaining symmetry is the $d=6$ conformal
subgroup of the full superconformal group.
Furthermore, the
variables $Y_1,Y_2\rightarrow 0$ and the supersymmetric
polynomial $T_{mn}\rightarrow t_{mn}$ and since
conformal symmetry is still present, the superconformal partial wave
expansion of the superspace  operator $\cO$ reduces to
the (non-supersymmetric) conformal partial wave expansion of a
component field of $\cO$.
One finds that the superconformal partial wave corresponding to 
 $\cO^4_{\cR_{MN}}$ becomes the conformal partial wave of
 $\vf^{\bf 55}_{M+4,N+4}$ (an  operator in the $\bf 55$ representation
 of $USp(4)$, with Lorentz  spin $j=M-N$ and dilation weight
 $D=M+N+8$) and the superconformal partial wave corresponding to
 $\cO^2_{M,0}$ becomes the conformal partial wave of the component
 $\vf^{\bf 14}_{M+2,2}$ (see table~\ref{table1}).

Explicitly then, the superconformal partial wave expansion of the
operator $\cO^4_{\cR_{MN}}$
\be\la{O4CPW}
<TTTT>\sim A_{MN}\ (g_{12} g_{34})^2
(\sdet Z)^2\sum_{m,n}  C^4_{MN}(m,n) {T_{M+m,N+n}\over\l^3}
\ee
in this limit becomes
\bea
<TTTT>&\sim &A_{MN} \ (y_{13}^2)^4 (x_{12}^2)^{-4}(x_{34}^2)^{-4}
\sum_{m,n}  C^4_{MN}(m,n) {t_{M+4+m,N+4+n} \over \l^3 }\\
&\sim& A_{MN}\ (y_{13}^2)^4 (x_{12}^2)^{-4}(x_{34}^2)^{-4}F_{M+4,N+4} 
\eea
see~\eqs{cpwabos}{FMN}. 
  This gives
\be
C^4_{MN}(m,n)=c_{M+4,N+4}(m,n)
\ee
where $c_{MN}(m,n)$ are the  coefficients defined in~\eq{FMN} for the
bosonic conformal partial wave expansion. 
Similarly
the superconformal partial wave expansion of the
operator $\cO^2_{\cR_{M0}}$ 
\be
<TTTT>\sim B_M\ (g_{12} g_{34})^2
\sdet Z\sum_{m,n}  C^2_M({m,n}) {T_{M+m,n}\over\l^3}\la{scpw2}
\ee
becomes
\bea
(y_{12})^{-2}(y_{34})^{-2}<TTTT>&\sim &B_M\ (y_{13})^4 (x_{12})^{-8}(x_{34})^{-8}
\sum_{m,n}  C^2_{M}(m,n) {t_{M+2+m,2+n} \over \l^3 }\\
&\sim& B_M\ (y_{13})^8 (x_{12})^{-8}(x_{34})^{-8}F_{M+2,2} 
\eea
where $F_{MN}$ is given in~\eq{FMN} giving
\be
C^2_{M}(m,n) =c_{M+2,2}(m,n).\la{scpwc}
\ee
In this way we have found the
complete superconformal partial wave expansion for $<TTTT>$.  

Note that the coefficients $A_{MN}$ and $B_{M}$ give  the following
combination of the three-point function coefficient and the two-point
function coefficient as defined in~\eqs{3pt}{s2pnt}
\be
{(A_{TT\cO})^2\over C_{\cO\cO}}.
\ee

It is possible to write the conformal partial wave expansion in the
form~\eq{Sbas} in terms of the functions $G,f,g$ defined
in~\eq{invform}.
For example, by inserting~\eq{scpwc} into~\eq{scpw2} and comparing
with~\eq{4ptTs} we find that the operator $\cO^2_{M0}$ contributes $G=f=0$ and
\be
\cO^2_{M0} \quad \rightarrow \quad 
g(X)\sim{X^{M+1}F_{21}(M+2,M+1;2M+4;X)}:=g_M(X)\la{gM}
\ee
where $F_{21}$ is a hypergeometric function.

\section{Superconformal partial wave analysis of the free and large N
  four point function.}\la{sec:cpwa}

\subsection{The free four-point
  function.} \la{freecpw}

The free four-point function has the following form:
\be \ba{rl}
<TTTT>= g_{13}^2 g_{24}^2& \left( A(1+\sdet (Z)^{-2} + \sd(1-Z)^{-2})\right.\\
     & +B\left.(\sdet (Z)^{-1} + \sdet (1-Z)^{-1} + \sdet (Z)^{-1}\sdet (1-Z)^{-1})\right).
\la{free4pt} \ea \ee
One can expand the superdeterminants in terms of Schur polynomials and
thus write this in the form~\eq{invform}:
\bea
<TTTT>&=& \l^{-4}g_{13}^2g_{24}^2 \cF[F]\\
\la{90}F(X_1,X_2)&=&A\left(1+{1\over
      X_1^2X_2^2}+{1\over(1-X_1)^2(1-X_2)^2}\right)\\\nonumber
  &+&B\left( {1\over X_1X_2}+{1\over(1-X_1)(1-X_2)}+{1\over X_1X_2(1-X_1)(1-X_2)}\right)\la{F0}
\eea

Decomposing $F$ into $G,f,g,A$ according
to~\eq{invform},
we obtain
\bea
G(X_1,X_2)&=&-{A(X_1-X_2)\over3(1-X_1)^3(1-X_2)^3}\\
f(X)&=&A\left(X+{1\over3(1-X)^3}\right)+B{X\over 1-X}\\
g(X)&=&B\left(X+{X\over 1-X}\right)\la{g0}\\
A&=&A
\eea

We can now expand these functions in terms of the conformal partial
waves calculated in the previous section. 
Using a computer one can check that
\be
g(X)=B(X+X/(1-X))=\sum_{j=0}^{\infty}B_{M}g_M(X)
\ee
where
\be 
 B_M=\left\{ \ba{ll}B\,\frac{\left(  M+2 \right)! \,{\left(  M \right) !}}
  {\left( 2\,M+1 \right) !}\quad &M \rm{even}\\
0&M \rm{odd} \ea \right.
\ee
 
and $g_M(X)$ are the conformal partial waves found in~\eq{gM}. 

The operator $\cO^4_{M\,0}$ contributes $g=0$ and
\be
f(X)\sim {\rm{d}^2 \over \rm{d
    x}^2}\left({X^{M+3}F_{21}(M+1,M+4,2M+8,X)\over(M+2)(M+3) }\right):=f_{M}(X).
\ee
In the free theory
\be
f(X)=A\left(X+{1\over 3(1-X)^3}-{1\over 3}\right)+B{X\over
  1-X}=\sum_{j=0}^{\infty}A_{M0}f_{M}(X)
\ee
where
\be 
 A_{M0}=\left\{ \ba{ll} 
  \frac{A\, \,\left( 2 + M \right) \,
     \left( 3 + M \right) !\,\left( 6 + M \right) !}
     {36\,\left( 5 + 2\,M \right) !}+\frac{B\,\left( 2 + M \right)^2 \,\left( M \right) !\,
     \left( 5 + M \right) !}{\left( 4 + M \right) \,
     \left( 5 + 2\,M \right) !}\quad &M \rm{even}\\
0&M \rm{odd}. \ea \right.
\la{AM0}\ee

Finally one may sum up the contribution (given in~\eq{O4CPW}) of all
operators $\cO^4_{MN}$ to $G(X_1,X_2)$ (with the help of~\eq{invform})
and compare with the free theory.
One has to solve the linear equations
\be
\sum_{N=0}^{\infty}\sum_{M=N}^{\infty}\sum_{m=0}^{\infty}\sum_{n=0}^{\infty}c_{M+4,N+4}(m,n)A_{MN}{t_{M+m,N+n}(X_1,X_2) \over
  X_1X_2(M+m+3)(N+n+1)}=\l^2G(X_1,X_2)
\ee
where $A_{M0}$ is given above. These equations can be solved order by
order. 
Again using a computer 
we can show
that the coefficients $A_{MN}$ are consistent with the formula

\be \la{AMN}
A_{MN}=\left\{ \ba{ll}\ba{l}{(M+3)!(M+4)!\over(2M+5)!}{(N+1)!(N+2)!\over(2N+1)!}(M-N+2)(M+N+5)\times\\
\left({A\over
      72}(M+N+6)+(-1)^N{B\over
      2}{1\over(M-N+1)(M-N+3)(M+N+4)}\right)\ea \qquad &M-N \rm{ even}\\[15pt]
\ba{l}0\ea&M-N \rm{ odd}\ea \right.\ee

Notice that in all of the above equations the coefficients
corresponding to conformal partial waves of operators with $M-N$ odd
vanish which is an important check on the calculations. Such
representations may not occur in the OPE of identical operators $T$.

\subsection{The large N AdS dual
  four-point function}\la{lncpw}

The four-point function calculated using supergravity on $AdS_7\xz S^4$ was
found in~\cite{AS}. It can be written in terms of a function of two
variables as:
\bea
<TTTT>&=& \l^{-4}g_{13}^2g_{24}^2 \cF[F]\\
F(X_1,X_2)&=&F_0(X_1,X_2)+\hat{F}(X_1,X_2)\\
\la{Fhat}\hat{F}(X_1,X_2)&=&-{B\l^2 \over 2 u
  v}(1-u\del_u)(1-v\del_v)(2+u\del_u+v\del_v)(1+u\del_u+v\del_v)
(uv\del_{uv})\F\\
\Phi&=&{1\over \l}\left(\ln \ X_1X_2 \ \ln {1-X_1\over
    1-X_2}+2\Li_2(X_1)-2\Li_2(X_2)\right)
\eea
where $F_0$ is the free theory  function~\eq{90} and $u:=X_1X_2\
v:=(1-X_1)(1-X_2)$.
The coefficients are given by
\be
A=1\qquad B={1\over N^3}
\ee
so we are here considering the large $N$ expansion of the theory
around the free 
theory with $A=1,\ B=0$ obtained for
$N\rightarrow \infty$ with first order corrections proportional to
$1/N^3$. The function $\hat F(X_1,X_2)$ has the form 
\be \hat
F(X_1,X_2)=F_c(X_1,X_2)+\rm{Log}(X_1X_2)F_d(X_1,X_2)
\ee
 where $F_c,F_d$ contain no Log
terms. The log term appears from the expansion of anomalous dimensions
depending on a parameter 
to first order in that parameter. The function $F_d$ therefore
contains information about the 
anomalous dimensions of operators in the large $N$ limit whilst the
function $F_c$ contains information about the renormalisation of the OPE
coefficients.

We  first perform a conformal partial wave analysis of $F_d$ to give the
anomalous dimensions. We find that under the decomposition of $F_d$ 
into $G,f,g,A$ according to~\eq{invform}, only the function $G$ is non-zero. 

By a similar procedure to that
used in the free 
theory case we find the following coefficients for $M-N$ even
\be
B_{MN}=\ba[t]{l}-{B\over 24} {(M+3)! (M+4)! \over (2M+5)!}{(N+2)!(N+4)!(N-1)_3
  \over (2N+1)!}
{1\over M-N+1}\left(1+{(N-2)(N+1) \over 2(M+N+4)(M-N+3)}\right)
\ea
\ee
The anomalous dimensions are given by dividing by the free theory
coefficients $A_{MN}$~\eq{AMN} with $B=0$
\be
\ba{l}
\c_{MN}={B_{MN}\over A_{MN}}=-{3B\over A} \left(1+{(N-2)(N+1) \over
    2(M+N+4)(M-N+3)}\right)  {(N-1)_6\over
  (M-N+1)(M-N+2)(M+N+5)(M+N+6)}.
\ea
\ee
Note that non-vanishing anomalous dimensions first appear for $N=2$
which correspond
to the first long operators in the theory. We may read off the anomalous
dimension of the operator $\cO^4_{22}$ (a long operator with lowest
component a scalar of dimension 8 in the free theory)
$\c_{22}=-24B/A$ in agreement with the results of~\cite{AS}.

Next we analyse the function $F_c$ to obtain information regarding
the OPE coefficients. Firstly we decompose $F_c$ into the functions
$G,f,g,A$ according to~\eq{invform}. 
We obtain 
\be g(X)=2 B X F_{21}(2,1;4;X)-B\left(X+{X\over 1-X}\right).
\ee
But $F_{21}(2,1;4;X)$ is the conformal partial wave of the
energy-momentum tensor (see~\eq{gM}) and $B\left(X+{X\over
    1-X}\right)$ is the contribution to $g$ of the free theory
(see~\eq{g0}) and so one finds that of all operators with charge 2, only the energy-momentum tensor
contributes to the four-point function in the large N limit. This fact
was previously observed in~\cite{AS}.

The analysis of $f$ gives the contribution of $\hat F$ to the
normalisation coefficients $A_{M0}$. These are given by
\be
\hat A_{M0}=B{(M+3)!^2(M+4)\over (2M+5)!(M+1)}\qquad M=0,2,4\dots
\ee
and should be added to the corresponding free theory expression~\eq{AM0}.
Analysis of $G$ leads to
\be
\hat A_{M1} =-3B {(M+3)!(M+4)!(3M^2+21M+28)\over M(M+2)(M+5)(2M+5)!}
\qquad M=1,3,5,\dots
\ee
which should also be added  to the corresponding free theory
expression in order to get the full coefficient. 

For $N\geq 2$ the analysis breaks down: the expressions get more and
more complicated,  and the coefficients no longer
vanish for  $M-N$ odd. This may be due to non-trivial mixing between
long operators with anomalous dimensions and the phenomenon also
occurs in the four dimensional $N=4$ SYM theory.

\section{Crossing symmetry}\la{crossing}

The four point function of four energy momentum multiplets has an
additional symmetry `crossing symmetry'. This simply states that the
four-point function is invariant under permutation of the insertion
points. It turns out that crossing symmetry has a very simple action
on the function $F(X_1,X_2)$.

Consider the four-point function~\eq{4ptcls}. Acting on the variable
$Z$ the permutations are generated by 
the two transformations  $Z\rightarrow 1-Z$ and $Z\rightarrow 1/Z$. 
Under $Z\rightarrow 1-Z$, $\cS$ is invariant as can
be seen from~\eq{Snice}, the differential function $\bf \D$ is invariant. This
transformation corresponds to $X_1\leftrightarrow X_3$ and so the
prefactor of the four-point function $(g_{13}g_{24})^2\l^{-4}$ is
invariant. So  we have that 
\be \la{cs1}
F(1-X_1,1-X_2)=F(X_1,X_2).
\ee

Under $Z\rightarrow Z^{-1}$, $\cS\rightarrow \cS/(X_1X_2Y_1Y_2)^2$, as can
be easily seen from~\eq{Snice}. The differential functional
is invariant and since this transformation corresponds to
$X_1\leftrightarrow X_4$ the prefactor is multiplied by $(Y_1Y_2)^2$.
Thus we find that 
\be\la{cs2}
F(1/X_1,1/X_2)=(X_1 X_2)^2\, F(X_1,X_2).
\ee
Note that one can easily see that the free theory function~\eq{90}
satisfies these symmetries. It is harder to see but also true that the
large N function $\hat F$ satisfies these symmetries.

\section{The four point function rewritten}\la{rewritten}

In order to compare our results with others it is useful to give 
the four point function in the following form:
\bea
<2222>&=&a_1 \xz g_{12}^2g_{34}^2+a_2 \xz g_{13}^2g_{24}^2+a_3\xz
g_{14}^2g_{23}^2 \ \ +\\\nonumber && b_1 \xz g_{13}g_{24}g_{23}g_{14}+b_2
\xz g_{12}g_{34}g_{14}g_{23}+ b_3 \xz  g_{12}g_{34}g_{13}g_{24}\\
&=& g_{13}^2g_{24}^2 \ba[t]{l}\left( a_1 {Y_1^2Y_2^2\over X_1^4X_2^4}
  + a_2 + 
  a_3 {(1-Y_1)^2(1-Y_2)^2\over (1-X_1)^4(1-X_2)^4} \right.\\
\left. + b_1{(1-Y_1)(1-Y_2)\over
    (1-X_1)^2(1-X_2)^2}+b_2{Y_1Y_2(1-Y_1)(1-Y_2)\over
    X_1^2X_2^2(1-X_1)^2(1-X_2)^2}+b_3{Y_1Y_2\over X_1^2X_2^2}\right)
\ea \eea
where $a_i,b_i$ are two-variable functions of $X_1,X_2$.

By writing $\cS$ in the following  form
\be \la{crossS}\ba{rcl}
S&=&X1 X2(1 - X1)(1 - X2) - Y1 Y2(X1 + X2)(1 - X1)(1 - X2) + \\
&&    Y1^2Y2^2(1 - X1)(1 - X2) + (1 - Y1)(1 - Y2)X1 X2(X1 + X2 - 2) + \\
&&    Y1 Y2(1 - Y1)(1 - Y2)(X1 + X2 - 2X1 X2) + (1 - Y1)^2(1 - Y2)^2X1 X2
\ea
\ee
in the expression for the four-point function~\eq{4ptcls} we can read off the
forms of the functions $a_i,b_i$
\footnotesize
\bea \la{a1}
\l^4a_1&=&u^4\left({\bf \D}(v F(X_1,X_2))+(1-X_1)^2F(X_1,X_1)+(1-X_2)^2F(X_2,X_2)\right)\\
\l^4a_2&=&{\bf \D}(uv F(X_1,X_2))+X_1^2(1-X_1)^2F(X_1,X_1)+X_2^2(1-X_2)^2F(X_2,X_2)\\
\l^4a_3&=&v^4\left({\bf \D}(u F(X_1,X_2))+(X_1)^2F(X_1,X_1)+(X_2)^2F(X_2,X_2)\right)\\
\l^4b_1&=&v^2\left({\bf \D}(u(X_1+X_2-2) F(X_1,X_2))-2X_1^2(1-X_1)F(X_1,X_1)-2X_2^2(1-X_2)F(X_2,X_2)\right)\\
\l^4b_2&=&u^2v^2\left({\bf \D}((X_1+X_2-2u) F(X_1,X_2))+2X_1(1-X_1)F(X_1,X_1)+2X_2(1-X_2)F(X_2,X_2)\right)\\
\l^4b_3&=&-u^2\left({\bf \D}(v(X_1^2-X_2^2)F(X_1,X_2))-2X_1(1-X_1)^2F(X_1,X_1)-2X_2(1-X_2)^2F(X_2,X_2)\right)\la{a6}
\eea

\normalsize

where
\be
\l:=X_1-X_2\qquad u:=X_1X_2\qquad v:=(1-X_1)(1-X_2).
\ee

We can now compare with the results of~\cite{AS} where the four-point
function was found using crossing symmetry. We find that the results
match except that in~\cite{AS} the additional terms depending on
$F(X_1,X_1)$ are absent.
The method employed their involved solving differential
equations for $a_1,a_3$ and $b_2$
coming from the superconformal Ward identities and then using crossing
symmetry to obtain the other functions $a_2,b_1,b_3$. The
equation for $a_1$ is solved in terms of a function $F_{AS}$ and
the solution has
the same form as the first term of equation~\eq{a1} (with $F$ replaced
by $F_{AS}$) but without the
additional two terms. Furthermore the differential  equations of~\cite{AS}
also demand 
that $F_{AS}(X,X)=0$.      
There is therefore a slight discrepancy between our results
and those of~\cite{AS} which disappears in the case that $F(X,X)=0$.
It does not seem to be possible to remove this discrepancy by
redefining $F$ since  the kernel of $\cF[F]$
is $F=0$ ie the four-point function is uniquely defined by $F$ (see 
the end of section~\ref{sexp}.) 
Note that on the other hand it is possible to rewrite the four-point
function  
such that the extra terms in $a_1$ disappear.
For example if we define a function $F_{AS}$ as
\be \la{fas}
F_{AS}(X_1,X_2)=F(X_1,X_2)-{k(X_1)-k(X_2)\over \l}
\qquad  k'(X)=F(X,X)
\ee
which clearly satisfies $F_{AS}(X,X)=0$ then by noting that ${\bf \D}(v
k(X_1)/\l)=(1-X_1)^2 k'(X)$ we find that~\eq{a1} becomes 
\be
\l^4 a_1=u^4 {\bf \D}(v F_{AS}).
\ee
The equation for $a_1$ has been simplified, the last two terms
of~\eq{a1} being absent. However the redefinition~\eq{fas} does not
in general remove the last two terms from all of
equations~\eqss{a1}{a6}. In fact it removes these additional terms
from $a_1,a_3$ 
and $b_2$ but at the cost of adding additional complicated $k$
dependent terms to
the other three coefficient functions. The crossing symmetry
conditions~\eqs{cs1}{cs2} for $F_{AS}$ will be 
more complicated  than they were for $F$ and will involve
the function $k(X)$ as 
well as $F_{AS}$ (but they will reproduce the crossing symmetry
relations~\eqs{cs1}{cs2} when $k'(X)=0$.)
In general then it seems one can only completely remove the additional
terms  in the case that $F(X,X)=0$. 
The function leading to the large $N$ four-point function $\hat
F$~\eq{Fhat} does satisfy $\hat F(X,X)=0$ so the results
of~\cite{AS} 
concerning the large $N$  four-point function are not affected by this
discussion.

\section{Discussion of protected operators in the $(2,0)$ theory}\la{protected}

In $N=4$ super Yang-Mills there are operators
which were originally assumed to be unprotected from renormalisation,
but which 
were discovered to have vanishing anomalous dimensions using
AdS/CFT. As 
representations in the free theory these operators lie at the
threshold of the continuous series a)~\eq{eq:ubds}. There
are in fact two types of operator which lie at this threshold,
ones like the Konishi operator which develop anomalous dimensions,
and other ones, which don't. In~\cite{Eden:2001wg} it was proved that
operators 
which occur in the OPE of two half-BPS operators and which saturate
the series a) bound are protected. 
The three-point function of two half-BPS operators and the
operator in question was examined and it was shown that an anomalous
dimension for the third operator would give a 3-point function,
incompatible with the superconformal symmetry.

In~\cite{ops} this phenomenon was
explained by a completely different and very simple argument, making use of the
classical interacting theory. Operators 
which are defined in terms of the chiral primary operators and lie
at the threshold of the unitary bound are short supermultiplets in the
classical interacting theory and
can not become long through the process of quantisation\footnote{There
  do exist apparently   short operators in the classical theory which
  develop anomalous dimensions (for example the quarter BPS descendant
  of the Konishi operator) but this is achieved because they can be
  written as 
 descendants of a long operator. 
This can not happen for
operators defined in terms of the half-BPS operators.}. Since
superconformal representation theory tells us that the operator with
an anomalous dimension must be a long supermultiplet we conclude that
all operators defined in terms of half-BPS operators and which
saturate the unitary bound are protected.
Those which can't be defined this way (such as Konishi)
are long and hence unprotected. In terms of the AdS/CFT
correspondence, the protected operators correspond to
multi-particle supergravity states and the unprotected operators
to string states.

Note that this second classification is much more general as one is
not restricted to operators obtained in the OPE of two half-BPS
operators. 

 The obvious question arises as to whether there is a similar
 phenomenon in six dimensions. In this case there is no known
 classical interacting theory so the general arguments of~\cite{ops}
 can not be applied. Furthermore none of the operators occurring
 in the OPE of two CPOs lies at the threshold of the series a) so the second
 method can not be straightforwardly applied either.
 However, one may try to generalise the argument from three-point
 function selection rules to analyse the three-point function of
 more complicated operators and find protected operators in this
 way. We shall illustrate this with an example below.
 
 Firstly however we consider the simple case of the three-point function
 of two CPOs 
 and an arbitrary third operator. This case was first considered
 in~\cite{Eden:2001wg}
 using a different method.

 We have
  \be
 \cT^{\unA}<\cO_{\unA}^Q \cO^p \cO^q> \sim  g_{12}^{\half (Q+p-q)} g_{12}^{\half (Q+q-p)}
 g_{12}^{\half (q+p-Q)} (X_{123}^{-1})_{\unA}\cT^{\unA}.
 \ee
 The first thing to notice is that on the right hand side the
 representation 
 $\cR$ carried by $\cT$ must be given by the Young tableau
 with the quantum numbers of $\cR$ as follows (see~\eq{YTcomp}):
 $n_1=n_3=0$ and $a_1, b$ 
 are even. This is because it is made from the tensor product of one
 object with two antisymmetric indices. Note that in this section we
 are using Young tableau in the form of~\eq{YTcomp} rather than the
 alternative form of~\eq{YTcomp2}.
 We must also consider restrictions due to analyticity of the left hand
 side. The largest poles in the variables $y_{12}$ and $y_{23}$ 
 and the restrictions they give are
 \be
\ba{lcl}
 (y_{12}^2)^{\half (Q+p-q)} (y_{12}^{-2})^{b \over 2} (y_{12}^{-1})^{a_1 \over
 2}
 & 
 \Rightarrow
& Q+p-q \geq a_1 + b  \\
(y_{13}^2)^{\half (Q+q-p)} (y_{13}^{-2})^{b \over 2} 
(y_{13}^{-1})^{a_1 \over 2} 
&\Rightarrow&Q+q-p \geq a_1 + b\\
(y_{23}^2)^{\half(p+q-Q)} (y_{23}^2)^{(b-4) \over 2}& \Rightarrow &
p+q-Q+b\qquad \rm{if } \ b\geq 4\\
(y_{23}^2)^{\half(p+q-Q)} &\Rightarrow&  p+q-Q \geq 0 
\qquad \rm{if }\ b=0,2 
\ea
\ee

 These results are consistent with the results of~\cite{Eden:2001wg}. Note that
 since $b$ 
 is even we can have no representations at the threshold of the series a)
 bounds.

 In order to obtain such operators we consider the three-point function of
 one half-BPS operator, $\cO^q$, one other series d) operator with two
 antisymmetric superindices, $\cO^p_{[AB]}$, and a third
 operator which for simplicity we choose to carry $SL(4|2)$
 representation given by the Dynkin indices
 $\cR=[000(3+2\c)a_1]$ (for the related Young tableau see~\eq{YTcomp}). 
 From the general three-point function formula~\eq{g3pt} we find
 \bea
 \cT^{\unB} \cT^{A_1A_2}<\cO_{\unB}^Q \cO^p_{A_1A_2} \cO^q> &\sim&
 \ba[t]{l}g_{12}^{\half (Q+p-q)} g_{13}^{\half (Q+q-p)} 
 g_{23}^{\half (q+p-Q)}  \xz \\ (X_{12}^{-1})_{A_1 C_1} (X_{12}^{-1})_{A_2 C_2}
 t(X_{123})_{\unB}^{C_1C_2} \cT^{\unC} \cT^{A_1A_2}.
\ea \\
t(X_{123})_{\unB}^{C_1C_2}&=&a
(X_{123}^{-1})^n_{\unB}(X_{123})^{A_1A_2}+b
(X_{123}^{-1})^n_{\unB_3}\d_{B_1}^{A_1}\d_{B_2}^{A_2} 
 \eea
where $n=4\c+6+a_1$ is the number of boxes in the Young tableau for
$\cR$ and $a,b$ are arbitrary coefficients. The indices $C_3$ are
forced into the subrepresentation $\cR'= 
[000(2+2\c)a_1]$.  
 Again $a_1$ is forced to be even.

 The lower bound on $Q$ comes from examining the pole structure  in
 $y_{23}$. There are potential poles in the propogator term that can
 be potentially cancelled by zeros
 in $X_{123}^{-1}=-X_{13}^{-1}X_{32}X_{21}^{-1}$. When $a=0$ we find
 the highest pole in $y_{23}$ is
  \be \ba{lcl}
 {\ba{l}(y_{23}^2)^{\half(p+q-Q)} (y_{23}^2)^{\c-1}\\
 (y_{23}^2)^{\half(p+q-Q)} \ea } 
 & {\ba{c} \Rightarrow\\ \Rightarrow 
 \ea}
 & {\ba{ll} Q-2\c \leq p+q-2 \qquad & \c \geq 1\\
 Q\leq p+q \qquad &\c=0 \ea } \ea
 \ee
 When $b=0$ on the other hand  we have an extra pole in $y_{23}$ from the term
 $X_{123}$. Therefore for this term to be analytic the charge $Q$ is
 even more restricted and we require  $Q\leq p+q-2$ even for $\c=0$. 
 Now the
 number $Q-2\c$ is independent of the anomalous dimension
 $\c$ and so an operator with $Q=p+q, \c=0$ can not develop an anomalous
 dimension.

Having shown that certain operators on the unitary bound are protected
we now discuss how other operators also saturating the bound might
develop anomalous dimensions.
In free $N=4$ SYM, an operator saturating the unitary bounds is short
and so in order for it to develop an
anomalous dimension it must become long by combining  with other
operators\cite{doanom,u1y}. The same is true in the present case and
analytic 
superspace provides a simple way of seeing which operators can combine
in this way. One simply needs to look at the corresponding isotropy
groups and which representations can combine to form long
representations. A nice way to find this is by considering the limit
as one lets 
 $b->3$ in the representation of the isotropy group for the long field
 in question. One gets different answers depending on which
 representation for the Young tableau you take (see
 section~\ref{sindices}). These two answers give the two short
 operators which make up the long operator.

The simplest example  is the superfield with Dynkin labels $[000b00]$
which has lowest component a scalar $USp(4)$ singlet operator of
dimension $2b$. As we let $b->3$, using the standard  Young tableau
of~\eq{YTcomp} and letting $b\rightarrow 3$ we arrive at a tableau with 2
columns and three rows corresponding to the operator lying on the
threshold of the unitary bounds. Using the alternative description
of the Young tableau however~\eq{YTcomp2} we get a tableau with one
column and  four rows. This corresponds to an operator with Dynkin
labels $[000040]$ which has lowest component a scalar in the $Usp(4)$
{\bf 35} representation and has  dimension 8. It lies in the series d)
series in the free theory but will not be protected from renormalisation.

More generally we have that the limit as $b\rightarrow 3$ of the
representation with Dynkin labels $[n_1n_2n_3xa_1a_2]$ (where
$x=b+n_1+n_2+n_3\rightarrow 3+n_1+n_2+n_3$) will split into two short
representations as follows:  
\bea
 \left[n_1
n_2 n_3 x a_1 
a_2\right] &\oplus& \left[(n_1-1) n_2 n_3 (x-1) (a_1+1) a_2\right]\ n_1 \geq 1\\
\left[0 n_2 n_3 x a_1
a_2\right] &\oplus& \left[0 (n_2-1) n_3 (x-2) a_1+1 a_2\right]
\  n_1=0\ n_2\geq 1\\
 \left[0 0 n_3 x a_1
a_2\right] &\oplus& \left[0 0 (n_3-1) n_3 (a_1+1) a_2\right]\qquad
n_1=n_2=0 \ n_3\geq1\\
 \left[0 0 0 3 a_1
a_2\right] &\oplus& \left[0 0  0 0 (a_1+4) a_2\right]\qquad
\qquad n_1=n_2=n_3=0.
\eea

Note that in SYM one has both primed and unprimed indices and
therefore in general one has four short operators combining to form a
long operator (for example the Konishi operator.) This is because we
have
two operators for each index type. Here there is only one index type
and hence  only two operators combine to form a long operator.

\section{Conclusion}

We have introduced the study of superconformal theories in six
dimensions using the analytic superspace formalism. Analytic
superspace is particularly suited to this task as it keeps all the
superconformal symmetry manifest from the beginning and has a similar
structure to ordinary Minkowski space so techniques can be
readily adapted from that context. In particular we
have shown how to find all correlation functions in the theory on
analytic superspace. 

We examined in detail the four-point
function of four energy-momentum multiplets. We found that there were
two different ways of
expanding the four-point function both of which were useful. The Schur
 polynomial expansion was useful in order to find a nice form for the
 four-point function in terms of a single function of two invariants.
 Another basis, $T_{\cR}$, which is a supersymmetric generalisation of
 the Jack polynomials introduced in~\cite{do6d}, was found to be useful
 in order to find the partial wave expansion. We then  performed a complete
conformal partial wave analysis of the free theory and the supergravity
dual theory, in particular giving the
anomalous dimensions of all operators in the
supergravity dual theory. 

We confirm that, as pointed out in~\cite{AS}, the free theory appears
to be disconnected from the supergravity dual theory since all
operators with charge 2 (other than the energy-momentum multiplet $T$)
occurring in the OPE of two Ts have
disappeared from the spectrum. This also 
happens in $N=4$ SYM where the operators are conjectured to  acquire
infinite anomalous dimensions in this limit, but here the operators
are protected by the superconformal unitary bounds (and it does not
seem to be possible to combine them with other operators to form long
operators as happens for example in the case of certain quarter BPS
operators in $N=4$ SYM~\cite{doqbps,dhhr}. 

In~\cite{AS} it was also noted that the function  $F$ describing  the
four-point functions in
the supergravity dual theory can be written in
the form $F=F_0+D F_{SYM}$ where $D$ is a differential operator and
$F_{SYM}$ is the corresponding function describing the four-point
function of energy-momentum multiplets in $N=4$ SYM at large N, and
large 't Hooft coupling $\l$. 
Since in the four-dimensional theory there  should  be a
smooth deformation connecting the free theory and the large $N$ theory
 provided by $F_{SYM}(\l)$ where $F_{SYM}(0)=0$ and  $F_{SYM}(\infty)=
 F_{SYM}$,  the interesting suggestion was made that one consider the
 object
\be
F=F_0+D 
 F_{SYM}(\l)\la{deform}
\ee which should provide a similar 
deformation for the six-dimensional theory.
It turns out however that on analysing the one loop four-point
function derived from this 
deformation using the CPWA one finds that the (protected) charge 2
operators attain 
anomalous dimensions in conflict with unitary bounds. The resolution
of the apparent conflict between the existence of this deformation and
the disappearance of the charge 2 operators from the spectrum is that
the deformation~\eq{deform} does not respect unitarity for all values
of the coupling.

Finally we considered operators at the threshold of the unitary bound
a) and found examples of operators which must be protected. 
 Notice that the protected operators in question have their $Q$
 charge~\eq{Q} equal 
 to the sum of the $Q$-charges of the other two operators in the three
 point function. This
 suggests that this rule can be generalised. One conjectures for
 example  that any
 operator with charge $Q$ which lies in the OPE of two protected
 operators of charges $p$ and $q$ where $Q=p+q$ and which saturates
 the series a) unitary bounds is itself protected.

\section*{Acknowledgements}
We would like to thank P. Howe for many helpful discussions.

\appendix

\section{Properties of $\c$ matrices in six dimensions}\la{sec:gamma}
 
In six space-time dimensions one has $8\times 8$ gamma matrices,
$\Gamma^{a}$ satisfying the Clifford algebra  
$\Gamma^{a}\Gamma^{b}+\Gamma^{b}\Gamma^{a}=2\eta^{ab}$
where we choose a mostly minus space-time signature. The gamma
matrices can be chosen such that
 \begin{equation}
\Gamma^{a}=\left(\begin{array}{cc}
                 0 & (\gamma^{a})_{\a\b}\\
                (\gamma^{a})^{\a\b} & 0 \end{array}\right)
\end{equation}
where $(\gamma^{a})_{\alpha\beta}=\textstyle{\frac{1}{2}}
\epsilon_{\alpha\beta\gamma\delta}(\gamma)^{a})^{\gamma\delta}$ and so
the Clifford algebra becomes
\be
(\c^a)_{\a\b}(\c^b)^{\b\c}+(\c^b)_{\a\b}(\c^a)^{\b\c}=2\d_{\a}^{\c}\eta^{ab}
\ee
which implies that
\bea
\e_{\a\b\c\d}x^{\a\b}x^{\c\e}&=&-2x^2\d_{\d}^{\e}\\
\e_{\a\b\c\d}x^{\a\b}x^{\c\d}&=&-8x^2.\la{x^2}
\eea
Another useful formula is
\be
\det(x^{\a\b})=(x^2)^2
\ee

\section{Analytic superspace in 6d} 

\subsection{Supercoset spaces of the superconformal group}

In~\cite{hartwell1} (see also~\cite{Luke}) it was shown that all complexified
superspaces of interest for 
studying globally supersymmetric theories in four-dimensions can be
viewed as supercoset spaces of the complexified superconformal
group. Then in~\cite{paris,sindices}
it was observed that these could all be represented by putting crosses on
a super Dynkin diagram from which one could also read off the
transformation properties of superfields. 

We follow the same route now to discuss six-dimensional superspaces. 
The complexified six-dimensional $(N,0)$ superconformal group is $Osp(8|2N)$
which has bosonic subgroup $SO(8) \times Sp(2N)$. The corresponding Lie
algebra, $\go \gs \gp (8|2N)$ can be represented as
the set of $(8|2N)\xz (8|2N)$ matrices satisfying

\be
\go \gs \gp (8|2N) =\{M | M J + JM^{ST} = 0 \}
\ee

where $M^{ST}$ denotes the supertanspose of $M$ and where

\be J= \left( \ba{cc|cc} 0_4 &1_4 & 0& 0 \\
                            1_4 &0_4 & 0& 0 \\
\hline                       0 & 0 &0_N & -1_N \\
                             0& 0  &1_N& 0_N \ea \right).
\la{J}\ee

 A general element of
$\go \gs \gp (8|2N)$ therefore has the form

\be \left( 
\ba{cc|cc}
A&B&\A&\B\\
C&-A^T&\C&\D\\
\hline
-\D^T&-\B^T&E&F\\
\C^T&\A^T&G&-E^T \ea \right)
\qquad
B+B^T=C+C^T=F-F^T=G-G^T=0
\ee

Complexified super Minkowski space is an open subset of the coset
space $P \bsh Osp(8|2N)$ where $P$ is the parabolic subgroup given by matrices
of the form

\be
P=\left\{ \left( \ba{cc|c} \bt & 0 & 0 \\
                           \bt & \bt&\bt\\
                    \hline \bt & 0 & \bt \ea \right) \right\}.
\ee

where bullets denote non-zero elements.
Minkowski superspace has coordinates $(x^{\a \b}, \th^{\a i})$
where $ \a,\b=1,\dots 4;
i=1 \dots 2N$ and $x^{\a \b}=-x^{\b\a}$. A coset representative is given by

\be
s(x,\th)= \left( \ba{cc|cc} 1 & x^{\a \b} & \th^{\a s} & \th^{\a s'} \\
                            0 & -1        & 0          & 0           \\
                 \hline     0 & \th_{r}{}^{ \b}& 1          & 0           \\
                            0 & \th_{r'}{}^{\b}& 0          & -1  \ea \right).
\ee

where $\th^{\a}_i=\eta_{ij} \th^{\a j}$ and where we have split the
indices $i=(r,r')$ with $r\in\{1,2\}$, $r'\in\{3,4\}$.

The super Dynkin diagram for $Osp(8|2N)$ in the form compatible with super
Minkowski space is given by\footnote{Unlike for the purely bosonic case, there
  can be different super Dynkin diagrams for the same supergroup. Different
  Dynkin diagrams lead to do different choices of simple roots, and hence to
  different possible parabolic subgroups. The
  Dynkin diagram here leads to the parabolic subgroup $P$ for super Minkowski
  space.}

\be
\begin{picture}(140,20)
\unitlength=1pt
\put(10,10){$\bt$}\put(30,10){$\bt$}\put(50,10){$\bt$}
\put(70,10){$\circ$}\put(90,10){$\bt$}\put(110,10){$\bt$}
\put(116,10){$\cdots$}\put(130,10){$\bt$}\put(150,10){$\bt$}
\put(12,13){\line(1,0){59}}\put(74,13){\line(1,0){37}}
\put(132,12){\line(1,0){20}}\put(132,14){\line(1,0){20}}
\put(90,10){$\overbrace{\phantom{a \hspace{55pt} a}}^N$}
\end{picture}
\ee

Then all superspaces of interest can be represented by putting crosses on the
nodes of this diagram. For instance, super Minkowski space is specified by
putting a single cross through the odd node (as in four dimensions).

In order to more easily see what the parabolic subgroups look like it is
convenient to change the basis to a more convenient form as follows

\be \left( \ba{c}
  v^{\a}\\
       v_{\a}\\
  \hline  v^{a}\\
          v_{a} \ea \right) \rightarrow
   \left( \ba{c} v^{\a}\\
       \hline  v^{a}\\
          v_{\tilde{a}} \\
\hline   v_{\a} \ea \right) \la{basis}
\ee

where $v_{\tilde{k}}=v_{(N-k)}$.
In this basis the parabolic subalgebras are simply given by block lower
triangular matrices.
For instance, for super Minkowski space we have

\be
P=\left\{ \left( \ba{c|c|c} \bt_4 & & \\
  \hline                   \bt &\bt_{2N}&\\
  \hline                   \bt & \bt&\bt_4 \ea \right) \right\}
\ee

where the subscripts indicate square matrices with the given size.

More generally, the Levi subgroup specified by putting crosses through
the nodes $(k_1, k_2, \dots k_{p-1})$, with $(k_{p-1} < N+4)$, of the Dynkin 
diagram is

\be
L=\left\{ \left( \ba{cccccccc}
\bt_{l_1}&&&&&&&\\
 & \bt_{l_2}&&&&&&\\
 &&\ddots&&&&&\\
 && & \bt_{l_p}&\bt_{l_p}&&&\\
 && & \bt_{l_p}&\bt_{l_p}&&&\\
 && &      &    &\bt_{l_{p-1}}&&\\
 && &      &    &       & \ddots&\\
 && &      &    &       &   & \bt_{l_1}
\ea \right) \right\}
\ee

where $l_i=k_i-k_{i-1}$ and we define $k_0=0, k_{p}=N+4$. The corresponding parabolic subgroup is
just the union of $L$ with the set of lower triangular matrices.
If there is a cross through the final node also, so the nodes $(k_1, k_2,
\dots k_{p-1}, N+4)$ are crossed through, then the Levi subgroup has the form

\be
L=\left\{ \left( \ba{cccccccc}
\bt_{l_1}&&&&&&&\\
 & \bt_{l_2}&&&&&&\\
 &&\ddots&&&&&\\
 && & \bt_{l_p}&         &&&\\
 && &          &\bt_{l_p}&&&\\
 && &      &    &\bt_{l_{p-1}}&&\\
 && &      &    &       & \ddots&\\
 && &      &    &       &   & \bt_{l_1}
\ea \right) \right\}
\ee

and again, the parabolic subgroup consists of the union of this set of
 matrices, with the
lower triangular matrices.

\subsection{Harmonic and analytic superspaces}\la{has}

We will be interested in Harmonic superspace and their related analytic
superspaces. Harmonic superspaces have super Dynkin diagrams with a cross
through the odd node and any number of further crosses through the nodes to
the right of the odd node. The related analytic superspaces have the same
Dynkin diagram, but with the odd node no longer crossed through.

In particular we can define $(N,p)$ harmonic superspaces by the Dynkin diagram
with just two crossed through nodes, the odd node and the $p$th internal
node.

\be
\begin{picture}(140,20)
\unitlength=1pt
\put(10,10){$\bt$}\put(30,10){$\bt$}\put(50,10){$\bt$}
\put(70,10){$\cxz$}\put(90,10){$\bt$}
\put(96,10){$\cdots$}\put(110,10){$\btx$}\put(116,10){$\cdots$}
\put(130,10){$\bt$}\put(150,10){$\bt$}
\put(12,13){\line(1,0){59}}\put(74,13){\line(1,0){17}}
\put(132,12){\line(1,0){20}}\put(132,14){\line(1,0){20}}
\put(90,10){$\overbrace{\phantom{a \hspace{15pt} a}}^p$}
\end{picture}
\ee

To find the coordinates of these spaces we split the internal index as
follows $a=(r,r'), r=1\dots p, r'= p+1 ,\dots N$, then the coordinates of
$(N,p)$ harmonic superspace are

\be
(x^{\a \b}, \th^{\a}_b, \th^{\a b}, y^r_{s'},y^{rs'},y^{rs})
\ee

and one can see that this space has the form Minkowski space times an internal
manifold.

The related analytic superspace has the Dynkin diagram

\be
\begin{picture}(140,20)
\unitlength=1pt
\put(10,10){$\bt$}\put(30,10){$\bt$}\put(50,10){$\bt$}
\put(70,10){$\circ$}\put(90,10){$\bt$}
\put(96,10){$\cdots$}\put(110,10){$\btx$}\put(116,10){$\cdots$}
\put(130,10){$\bt$}\put(150,10){$\bt$}
\put(12,13){\line(1,0){59}}\put(74,13){\line(1,0){17}}
\put(132,12){\line(1,0){20}}\put(132,14){\line(1,0){20}}
\put(90,10){$\overbrace{\phantom{a \hspace{15pt} a}}^p$}
\end{picture}
\ee

This has coordinates

\be
(x^{\a \b}, \th^{\a}_{s'}, \th^{\a a},y^r_{s'},y^{rs'},y^{rs})
\ee

and we see that this has the form of Harmonic superspace but with fewer odd
coordinates. In fact there are only $4(2N-p)$ odd coordinates instead of the
maximal $8N$.

More general $(N,p)$ harmonic superspaces can also be defined which have the
same two crosses through the odd node and the $p$th internal node, but with
further crosses through nodes to the right of the $p$th internal node. The
corresponding analytic superspaces will still have $4(2N-p)$ odd coordinates, but
will have a different internal space.

Of most interest for us will be $(N,N)$ analytic superspace which just has one
cross through the final $N$th node

\be
\begin{picture}(140,20)
\unitlength=1pt
\put(10,10){$\bt$}\put(30,10){$\bt$}\put(50,10){$\bt$}
\put(70,10){$\circ$}\put(90,10){$\bt$}\put(110,10){$\bt$}
\put(116,10){$\cdots$}\put(130,10){$\bt$}\put(150,10){$\btx$}
\put(12,13){\line(1,0){59}}\put(74,13){\line(1,0){37}}
\put(132,12){\line(1,0){20}}\put(132,14){\line(1,0){20}}
\put(90,10){$\overbrace{\phantom{a \hspace{55pt} a}}^N$}
\end{picture}
\ee

This has coordinates $(x^{\a \b},\l^{\a b}, y^{a b})$ and we see that there are
only $4N$ odd coordinates.

If we define the superindex $A=(\a, a)$, then the
parabolic subalgebra for
$(N,N)$ harmonic superspace is given by

\be
\gp=\left\{ \left( \ba{cc} -A^A{}_B &0 \\
                         - C_{AB}   &D_{A}{}^B \ea \right) \right\}
\ee

where

\be
D_{A}{}^{B}= (-1)^{A(A+B)} A^{B}{}_A \la{D=}
\ee
 ie

\be 
D=A^{ST}
\ee

and all three entries are $(4|N)\xz (4|N)$ matrices. As in the four
dimensional case we choose $\a$ to be even and $a$ to be odd.

A coset representative for
  the space is

\be \la{cosrep}
s(X)=\left( \ba{cc} 1 &X\\
                    0 &-1 \ea \right)
\ee

where the components of $X$ are

\be\la{scoords}
X=\left( \ba{c|c} x^{\a \b} & \l^{\a b} \\
       \hline         -\l^{a \b}& y^{ab} \ea \right).
\ee

Now $x^{\a \b} + x^{\b \a} =0$, $y^{a b} - y^{b a} = 0$ and so we see that
\be
 X^{AB}= -(-1)^{AB} X^{BA},
\ee

ie. $X$ is (generalised) antisymmetric.

Using standard coset space techniques one can then show that a general
infinitesimal conformal transformation
\be
\left(\ba{cc}
-A&B\\
-C&A^{sT}
\ea \right)
\ee
acting on the group leads to the following infinitesimal transformation
of the coordinates X
\be
\d X= B + AX + XA^{sT} +XCX. 
\ee
Here $A,B,C,X$ are all $(4|2)\xz (4|2)$ supermatrices, and $B$ and $C$ are
generalised anti-symmetric.

\subsection{Representations of the superconformal group} \la{sec:ubds}
Unitary representations of the real superconformal group in six dimensions
have were classified in~\cite{dobrev6d}.

They are given in terms of the labels,

\be [d; n_1,n_2,n_3;a_1,\dots,a_N]
\ee

where $d$ is the conformal weight, $n_1,n_2,n_3$ are the Dynkin labels
specifying the representation of the six-dimensional Lorentz group $SO(5,1)$
and $a_1,\dots a_N$ are Dynkin labels for the representation of the internal
group $Usp(2N)$. These are related to the super Dynkin coefficients of the
super conformal group as follows

\be
\begin{picture}(140,20)
\unitlength=1pt
\put(10,10){$\bt$}\put(30,10){$\bt$}\put(50,10){$\bt$}\put(70,10){$\circ$}
\put(90,10){$\bt$}\put(110,10){$\bt$}\put(116,10){$\cdots$}
\put(130,10){$\bt$}\put(150,10){$\bt$}
\put(12,13){\line(1,0){59}}\put(74,13){\line(1,0){37}}
\put(132,12){\line(1,0){20}}\put(132,14){\line(1,0){20}}
\put(10,20){\tiny $n_1$}\put(30,20){\tiny $n_2$}\put(50,20){\tiny
  $n_3$}\put(70,20){\tiny $x$}
\put(90,20){\tiny $a_1$}\put(110,20){\tiny $a_2$}\put(118,20){\tiny $\cdots$}
\put(128,20){\tiny $a_{N-1}$}\put(150,20){\tiny $a_N$}
\end{picture}
\ee

where

\be
x={d \over 2} + {1 \over 4}(n_1 +2n_2 +3n_3) - m_1
\ee

where $m_1= \sum_i a_i$.

The unitary representations fall into four series as follows
\be
\ba{lrcl}
a) \qquad& d &\geq& \half (3n_1 +2 n_2 +  n_3 ) + 2m_1 +6\\
b)       & d & = & \half ( n_3 + 2 n_2 )+ 2m_1 + 4, \quad n_1=0\\
c)       & d & = & \half  n_3 + 2m_1 + 2, \quad n_1=n_2=0\\
d)        & d & = & 2m_1, \quad n_1=n_2=n_3=0.
\ea \ee

In terms of super Dynkin labels these bounds become
\be \la{eq:ubds}
\ba{lrcl}
a) \qquad&  x& \geq& n_1 +n_2+n_3 +3\\
b) \qquad& x &  =  & n_2+n_3+ 2, \quad n_1=0\\
c) \qquad&x&   =   & n_3 +1, \quad n_1=n_2=0\\
d) \qquad&x&  =    &  n_1=n_2=n_3=0.\\
\ea \ee 

Notice that for the series $d)$ representations the first four super Dynkin
labels vanish.

All series d) representations can be given as ordinary superfields (ie without
superindices, although they may have internal indices) on $(N,1)$ analytic
superspace and therefore only depend on at most $4(2N-1)$ odd
coordinates. Series d) representations with
$a_1=\dots=a_{N-1}=0$ can be given as scalar fields in $(N,N)$ analytic
superspace and hence only depend on half the total number of odd
coordinates. These are the representations which are dual to Kaluza Klein
states in the AdS/CFT correspondence.

\subsection{Representations on analytic superspace: superindices}\la{sindices}

We wish to describe all unitary representations as superfields (possibly with
superindices) on $(N,N)$ analytic superspace. Superfields on $(N,N)$ analytic
superspace carry linear representations of $\gs \gl(4|N)\oplus \com = \gg
\gl(4|N)$. This linear representation can be read off from the super Dynkin
diagram, and for $x\in \int$ it can be obtained by tensoring together the
anti-fundamental representation (all unitary representations have downstairs
superindices). Tensor representations can be defined using Young
tableaux.

Abstractly, a representation carried by a superfield on $(N,N)$ analytic
superspace is given by the Dynkin diagram

\be
\begin{picture}(140,20)
\unitlength=1pt
\put(10,10){$\bt$}\put(30,10){$\bt$}\put(50,10){$\bt$}\put(70,10){$\circ$}
\put(90,10){$\bt$}\put(110,10){$\bt$}\put(116,10){$\cdots$}
\put(130,10){$\bt$}\put(150,10){$\btx$}
\put(12,13){\line(1,0){59}}\put(74,13){\line(1,0){37}}
\put(132,12){\line(1,0){20}}\put(132,14){\line(1,0){20}}
\put(10,20){\tiny $n_1$}\put(30,20){\tiny $n_2$}\put(50,20){\tiny
  $n_3$}\put(70,20){\tiny $x$}
\put(90,20){\tiny $a_1$}\put(110,20){\tiny $a_2$}\put(118,20){\tiny $\cdots$}
\put(128,20){\tiny $a_{N-1}$}\put(150,20){\tiny $a_N$}
\end{picture}
\ee

We read off the linear representation of the isotropy group $\gs
\gl(4|N)$ which the 
superfield carries. This is given by the Dynkin diagram

\be
\begin{picture}(140,20)
\unitlength=1pt
\put(10,10){$\bt$}\put(30,10){$\bt$}\put(50,10){$\bt$}\put(70,10){$\circ$}
\put(90,10){$\bt$}\put(110,10){$\bt$}\put(116,10){$\cdots$}
\put(130,10){$\bt$}
\put(12,13){\line(1,0){59}}\put(74,13){\line(1,0){37}}
\put(10,20){\tiny $n_1$}\put(30,20){\tiny $n_2$}\put(50,20){\tiny
  $n_3$}\put(70,20){\tiny $x$}
\put(90,20){\tiny $a_1$}\put(110,20){\tiny $a_2$}\put(118,20){\tiny $\cdots$}
\put(128,20){\tiny $a_{N-1}$}
\end{picture}
\ee

This is related to the following Young tableau

\be \setlength{\unitlength}{.25mm}
\begin{picture}(420,170)(0,-30)
\put(100,140){\framebox(180,20){}}
\put(100,120){\framebox(120,20){}}
\put(100,100){\framebox(60,20){}}
\put(20,-30){\framebox(20,190){}}
 \put(40,0){\framebox(20,160){}}
\put(60,30){\framebox(20,130){}}
 \put(80,60){\framebox(20,100){}}
\put(20,-20){$\left. \ba{c} \phantom .\\ \vspace{-20pt} \\
\phantom . \ea \right\}{}_{a_{N-1}}$} \put(60,40){$\left. \ba{c}
\phantom .\\ \vspace{-20pt} \\ \phantom . \ea
  \right\}{}_{a_{1}}$}
\put(85,105){${}_b$} \put(120,105){${}_{n_1}$}
 \put(160,120){$\underbrace{\hspace{45pt}}_{n_2}$}
 \put(220,140){$\underbrace{\hspace{45pt}}_{n_3}$}
\end{picture}\la{YTcomp}
 \ee

where $b=x-(n_1 +n_2 +n_3)$.

The three series of unitary bounds now come simply from demanding
that the Young tableaux have the correct shape.
 Clearly in the generic case we require that $b \geq 3$, but if
 $n_1=0$ we are allowed $b=2$, if $n_1=n_2=0$ we can have $b=1$
 and if $n_1=n_2=n_3=0$ we can have $b=0$, giving precisely the
 four series of unitary bounds above.

It will also be useful to define the quantum number $Q$

\be Q=(x+m_1 - n_1 -n_2-n_3)=(b+m_1) \la{Q}\ee

which in terms of the dilation weight $d$ is

\be
 Q= \half d - {1 \over 4} (3n_1 + 2n_2+n_3).
\ee

This is sometimes referred to as `twist' in the literature.

Note that if $b>0$ there is a Young tableau  related to~\eq{YTcomp} 
which corresponds to the same $\gs\gl(4|N)$ representation but different
$\gg\gl(4|N)$ representations.
This has the form
\be \setlength{\unitlength}{.25mm}
\begin{picture}(420,170)(0,-10)
\put(100,140){\framebox(180,20){}}
\put(100,120){\framebox(140,20){}}
\put(100,100){\framebox(100,20){}}
\put(100,80){\framebox(60,20){}}
\put(20,-30){\framebox(20,190){}}
 \put(40,0){\framebox(20,160){}}
\put(60,30){\framebox(20,130){}}
 \put(80,80){\framebox(20,80){}}
\put(20,-20){$\left. \ba{c} \phantom .\\ \vspace{-20pt} \\
\phantom . \ea \right\}{}_{a_{N-1}}$} \put(60,40){$\left. \ba{c}
\phantom .\\ \vspace{-20pt} \\ \phantom . \ea
  \right\}{}_{a_{1}}$}
 \put(120,85){${}_{b-4}$}
 \put(160,100){$\underbrace{\hspace{25pt}}_{n_1}$}
 \put(200,120){$\underbrace{\hspace{25pt}}_{n_2}$}
 \put(240,140){$\underbrace{\hspace{25pt}}_{n_3}$}
\end{picture}\la{YTcomp2}
 \ee
With this form of the Young tableau the charge Q is modified to
\be
Q=4+m_1.
\ee
We use the first form of the  Young tableau when discussing operators in
section~\ref{protected} whereas for the discussion of the four-point
function we use this second form.

\section{Superconformal invariants} \la{invariants}

The procedure for finding arbitrary four-point invariants in $N=4$ SYM
given in~\cite{u1y} may be readily adapted to the present case. We
sketch this here.

Using
similar arguments to those given for the case of the four-point
function around~\eq{dx213} one can reduce the problem of finding an
n-point function $F(X_1,X_2, \dots X_n)$ invariant under
superconformal transformations (acting on the $X$s as in \eq{dx}) to
that of finding a function $F(Z_1,Z_2,\dots Z_{n-3})$ invariant under
the adjoint representation of $OSp(4|N)$. 

One can then proceed in one
of two ways. We can form an invariant function as a polynomial in the
$(Z_i)^A{}_B$s and either take the superdeterminant of this or 
suitably contract all  the indices with $\d^A_B$s.~\footnote{Note
  that we may use any numerically invariant tensors we have at our
  disposal to construct invariants. In the present context we also have an
  invariant tensor $K^{AB}$ used in the definition of
  $OSp(4|N)$~\eq{osp}. We can not straightforwardly use $K$ however
to contract indices since it has two upstairs indices and any object
constructed from $Z$s has an equal number of upstairs and downstairs
indices. It may well, however,  be possible to construct invariants by
using $K$s and $\d$s in conjunction with an invariant $\cE$
tensor. Such an $\cE$ tensor was introduced in~\cite{u1y} in the
context  invariants in $N=4$ SYM and reflects the equivalence of
different tensors discussed in section~\ref{sindices}. In the present
context (for $N\neq4$) such a tensor would have different numbers of
upstairs and
downstairs indices (unlike for $N=4$ SYM) and so together with $K$
could lead to invariants. It would be
interesting to investigate this further.}
For example for the four-point function we have discussed expansions
of Z in terms of Schur polynomials and the polynomials $T{\cR}(Z)$
which are all polynomials in traces of powers of $Z$.

Another approach consists of systematically using up all remaining
symmetries and thereby reducing the number of components in the
$Z$s. The remaining components will be invariants\footnote{Strictly
  they are only invariant under transformations connected to the
  identity. There may still be global discrete transformations acting
  on these variables. For example in the $n=4$ case we have seen that
  four-point functions are invariant under $X_1\leftrightarrow X_2$}.

The first stage of this procedure consists of diagonalising $Z_1$
to the form of~\eq{diagz}. In the case $n=4$ we stop here but for
higher point functions we can go further. The residual
infinitesimal symmetry leaving $Z_1$ invariant has the form:
\be
A=\left(\ba{c|c}{\bf a}&0\\\hline 0&0\ea\right)
\qquad
{\bf a} = \left(\ba{cc}D_1 &D_2\\ D_3&-D_1\ea\right)\ee
where the $D_i$ are $2\xz2$ diagonal matrices. One then uses up as
much of this symmetry to obtain a specific form for $Z_2$ which may
still be
invariant under a smaller residual symmetry in which case one uses
this on $Z_3$ etc. When all the infinitesimal symmetry has been absorbed
we are necessarily left with invariants (up to discrete symmetries.)

\bibliographystyle{../bibliography/utcaps}
\bibliography{../bibliography/bibliography2}

\providecommand{\href}[2]{#2}\begingroup\raggedright\endgroup

 \ed